# Tunable Graphene Electronics with Local Ultrahigh Pressure

P. Ares[1*], M. Pisarra[2,3], P. Segovia[1,4], C. Díaz[2,4,5], F. Martín[2,3,4], E. G. Michel[1,4], F. Zamora[3,4,5,6], C. Gómez-Navarro[1,4], and J. Gómez-Herrero[1,4*]

[1] Departamento de Física de la Materia Condensada, Universidad Autónoma de Madrid, Madrid, E-28049, Spain
[2] Departamento de Química, Universidad Autónoma de Madrid, Madrid, E-28049, Spain
[3] Instituto Madrileño de Estudios Avanzados en Nanociencia (IMDEA-Nanociencia) Cantoblanco, Madrid, E-28049, Spain
[4] Condensed Matter Physics Center (IFIMAC), Universidad Autónoma de Madrid, Madrid, E-28049, Spain
5 Institute for Advanced Research in Chemical Sciences (IAdChem), Universidad Autónoma de Madrid, Madrid, E-28049, Spain
[6] Departamento de Química Inorgánica, Universidad Autónoma de Madrid, Madrid, E-28049, Spain

* E-mail: pablo.ares@uam.es, julio.gomez@uam.es

We achieve fine tuning of graphene effective doping by applying ultrahigh pressures (> 10 GPa) using Atomic Force Microscopy (AFM) diamond tips. Specific areas in graphene flakes are irreversibly flattened against a $SiO_2$ substrate. Our work represents the first demonstration of local creation of very stable effective p-doped graphene regions with nanometer precision, as unambiguously verified by a battery of techniques. Importantly, the doping strength depends monotonically on the applied pressure, allowing a controlled tuning of graphene electronics. Through this doping effect, ultrahigh pressure modifications include the possibility of selectively modifying graphene areas to improve their electrical contact with metal electrodes, as shown by Conductive AFM. Density Functional Theory calculations and experimental data suggest that this pressure level induces the onset of covalent bonding between graphene and the underlying $SiO_2$ substrate. Our work opens a convenient avenue to tuning the electronics of 2D materials and van der Waals heterostructures through pressure with nanometer resolution.

## 1. Introduction

Reaching ultrahigh pressures always implies an immense experimental challenge. The most common device employed is the diamond anvil cell,[1] where two opposing diamond polished tips compress a submillimeter-sized sample. Diamond anvil cells can typically reach pressures of 100-200 GPa, and maximum values as high as 750 GPa may be obtained by employing special cells.[2] The extraordinary strength, stiffness and high flexibility[3] of graphene make it an ideal material to withstand ultrahigh pressure.[4] Indeed, these outstanding mechanical properties have already proved to be a promising route to induce chemical reactions on molecules trapped by graphene.[5] In this work, we have taken advantage of these properties to perform controlled local modifications of graphene deposited on $SiO_2$ substrates. We achieve ultrahigh pressures by means of Atomic Force Microscopy (AFM), using cantilevers with diamond tips. This simple experimental set up, which can be considered as the nanotechnology version of a diamond anvil cell, has allowed us to reach pressures well above 40 GPa.[6]

Our results shed light on the adhesion of graphene on $SiO_2$/Si substrates, a paramount issue for both fundamental understanding and technological applications. By applying ultrahigh pressures, we have tuned the adsorption strength, an important factor controlling graphene electronic properties. [7] We have used this method to create selected areas in graphene on $SiO_2$/Si substrates where holes are the majority charge carriers (they present positive doping or, in other words, they are p-doped), as demonstrated by Raman spectroscopy, Kelvin Probe Force Microscopy (KPFM) and Scanning X-ray Photoelectron Microscopy (SPEM) characterizations. A direct application of this local tuning of graphene electronic properties is the creation of regions with low electrical contact resistance, highly relevant for graphene electronics. Density Functional Theory (DFT) calculations, in good agreement with the SPEM data, confirm that ultrahigh pressure induces the onset of covalent bonding between graphene and $SiO_2$, explaining, with a single hypothesis, both the new equilibrium distance and the effective p-doping. Our work presents the proof of concept of a novel procedure to locally tune the adhesion and electronic properties of graphene on technological substrates;

suggests a well-controlled platform to develop new ultrahigh pressure chemistry at the nanoscale; and opens a simple way to experimental tuning of the electronic properties of van der Waals heterostructures such as twisted bilayer graphene and related systems.[7c, 8]

## 2. Results

Achieving controlled local high pressures with AFM is technically remarkably simple.[6] Typical commercial diamond AFM tip radii are in the range of 10 - 100 nm. Thus the tip-sample contact region can be seen as a nano-anvil cell, where ultrahigh pressures can be readily achieved with forces in the µN range (see Experimental section).

We deposited graphene flakes by microexfoliation on 300 nm $SiO_2$/Si substrates. Inspection by optical microscopy allowed us to locate single layer graphene areas, which were later corroborated by Raman spectroscopy[9] (**Figure 1**). Figure 1a shows an optical microscopy image of a graphene representative flake with several monolayer terraces. Raman spectra acquired in the different regions of the flake corroborate the thickness of each of the terraces. Figure 1b shows a Raman spectrum acquired in a monolayer region. The sharp 2D peak shape and the ratio intensity between G and 2D peaks confirm that it corresponds to a single-layer terrace. Additionally, the absence of a D peak indicates that only a negligible number of defects is present on the graphene layer, as expected for flakes produced by microexfoliation.

The procedure to modify areas under ultrahigh pressures involves the following steps: firstly, we carry out a non-invasive gentle AFM image of the flake in dynamic mode. Then, we bring the tip into contact mode and apply a load corresponding to a given pressure on a selected area (see Experimental section). We scan the area twice under these ultrahigh-pressure conditions: typically, we scan the area in the slow scan direction from the top to the bottom, and after reaching it, from there to the top. Finally, we bring back the tip to dynamic mode and acquire a new topography image. Figure 1c shows the upper right corner region of the flake shown in Figure 1a, where we modified several 600 x 600 $nm^2$ areas in monolayer terraces at different pressures, starting from 13 GPa and up to 40 GPa. Figure S1 in the Supporting Information (SI) presents modified areas of different sizes and shapes, confirming the robustness of our procedure as we are not limited to small areas.

From Figure 1d it can be seen that, for pressures below 13 GPa, we did not modify the graphene area. From 16 to 25 GPa, the height in the topography images observed within the modified areas decreased by ~1 nm. For pressures in the range of 29 to 37 GPa, the modified areas' height was lowered by an additional ~0.3 nm, such that the maximum displacement was ~1.3 nm. Finally, for pressures higher than 40 GPa the graphene sheet broke and the tip induced irreversible damage in the underlying $SiO_2$. We performed similar modifications directly on the $SiO_2$ substrate, without graphene (see Figure S2). In this case, we did not modify the $SiO_2$ substrate up to pressure values of ~25 GPa (the tip swept some debris from the scanned area but its depth did not change). For higher pressures, it was clearly modified when compared to graphene-covered $SiO_2$, pointing out the abilities of graphene as a wear protection coating.[10] It is worth mentioning that the effect of the pressure-induced modifications in the graphene/$SiO_2$ sample were irreversible and stable over time. We found similar height profiles before and after keeping the sample under ambient conditions for four months (see Figure S3).

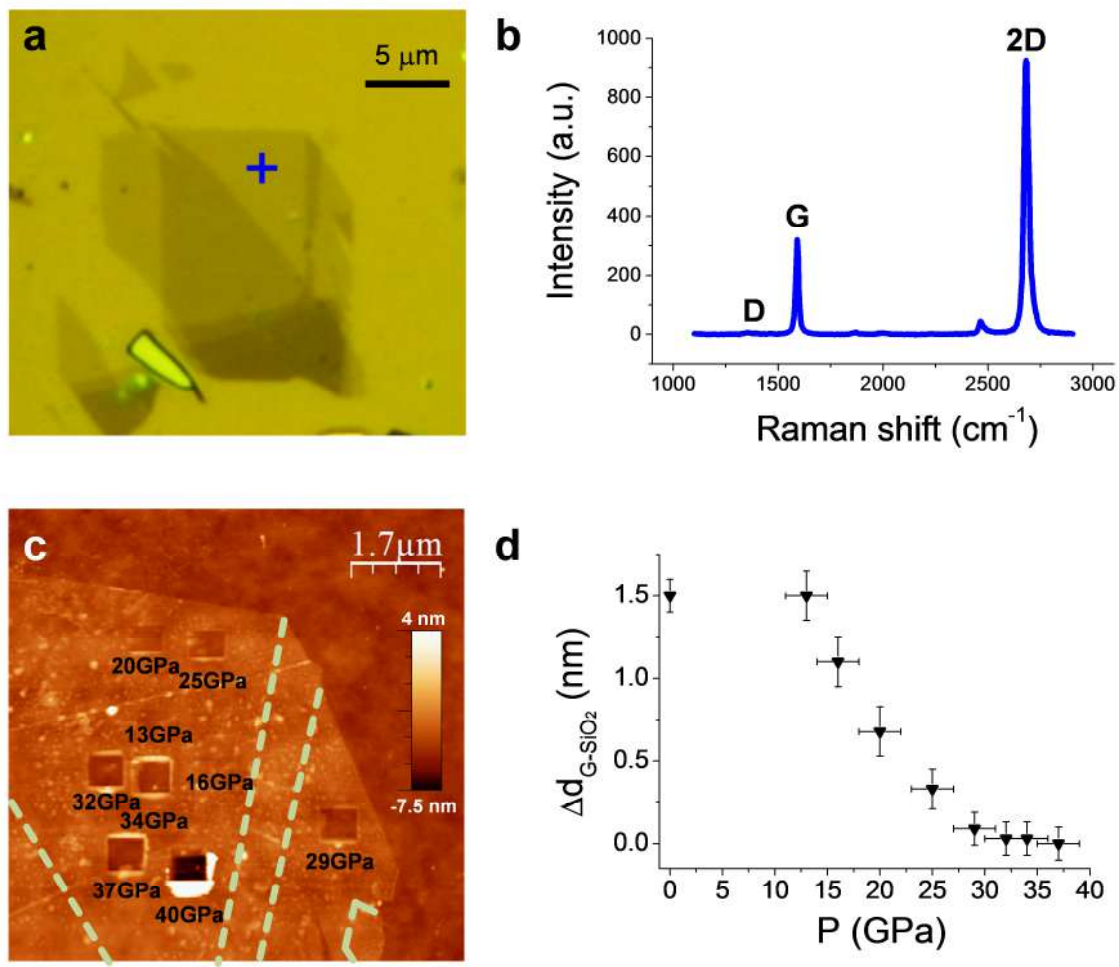

**Figure 1.** Characterization of pristine graphene on SiO$_2$ and ultrahigh pressure modifications. a) Optical image b) Raman spectrum acquired in the spot marked with a cross in a), showing the graphene characteristic peaks. c) AFM topographic image showing 600 x 600 nm$^2$ areas modified under different pressures. d) Graphene-SiO$_2$ distance variations as a function of the applied pressure. $\Delta d_{G-SiO_2}$ accounts for the difference in pristine graphene and modified regions surface heights. $\Delta d_{G-SiO_2} = 0$ has been chosen as the minimum graphene-SiO$_2$ distance (corresponding to the deepest modified graphene area before breakage).

In order to study the physical properties of the modified regions, we carried out an analysis using Raman spectroscopy and Kelvin Probe Force Microscopy (KPFM). In **Figure 2** we report the Raman spectra and KPFM profiles of the modified regions of Figure 1; further experimental results are provided in the SI (Figure S4-S6). Figure 2a shows the evolution of the G and 2D peaks within the modified areas up to 37 GPa: a clear non-rigid shift of the peaks and changes in width and intensity are observed. Interestingly, no D peak appeared upon ultrahigh pressure modifications (Figure S6), indicating that scanning at high pressure does not generate defects within the Raman sensitivity threshold, (typical maximum distance between defects of ~ 20 nm).[11] The evolution of the G and 2D peaks' line-shape and position, and the measured I(2D)/I(G) ratio (Figure S5),[12] indicate a p-doping effect, in very good agreement with the reported hole doping of graphene through gated transistor configurations.[12-13]

Figure 2b (see also Figure S4) presents KPFM profiles of the modified areas. KPFM is a standard technique used to characterize doping in two-dimensional materials (see Experimental section for details). A clear reduction of the Contact Potential Difference (CPD) with increasing pressure can be observed, indicating that the Fermi level shifts down with respect to the Dirac cone vertex as the applied pressure increases. This observation is compatible with a pressure-dependent hole doping due to a charge transfer effect,[14] and is also in excellent agreement with the Raman observations. We obtained additional insights about the charge transfer between graphene and substrate by analyzing the observed CPD. Its variation can be converted into a variation in the Fermi level using the following expression:[15]

$$\Delta E_F = e\Delta V_{CPD} \qquad (1)$$

where $e$ is the elementary charge. Given the Fermi level shift, the variation of the carrier concentration in the graphene modified areas can be estimated as:[16]

$$\Delta n = \frac{1}{\pi}\left(\frac{\Delta E_F}{\hbar v_F}\right)^2 \qquad (2)$$

where $\hbar$ is the Planck constant, and $v_F$ is the Fermi velocity ($|v_F| = 10^6$ m s$^{-1}$). Figure 2c displays a multiple axis chart where we plotted the I(2D)/I(G) ratio (red triangles, left axis), the graphene-SiO$_2$ distance variation (black inverted triangles, right black axis) and the Fermi level shift (blue crosses, right blue axis) as functions of the applied pressure. Figure 2d shows the variation of the Fermi level as a function of the variation in the graphene-substrate distance. Finally, Figure 2e presents the 2D/G intensity ratio as a function of the variation of the electron concentration. It can be observed that the 2D/G intensity ratio and the Fermi level shift present a similar dependence with increasing pressure, being both proportional to the graphene-SiO$_2$ distance variation within the modified areas. The results obtained are in good agreement with conventional graphene doping achieved using interface engineering,[16] chemical approaches,[17] thermal annealing and gas flow experiments,[7b] electrostatic gating,[12-14] or combinations of such techniques.[18] In our experiments, the pressure applied to graphene produces a tunable decrease in the graphene-substrate distance, and therefore an increase of the coupling to the substrate, resulting in a tunable doping level of selected areas with nanometer precision.

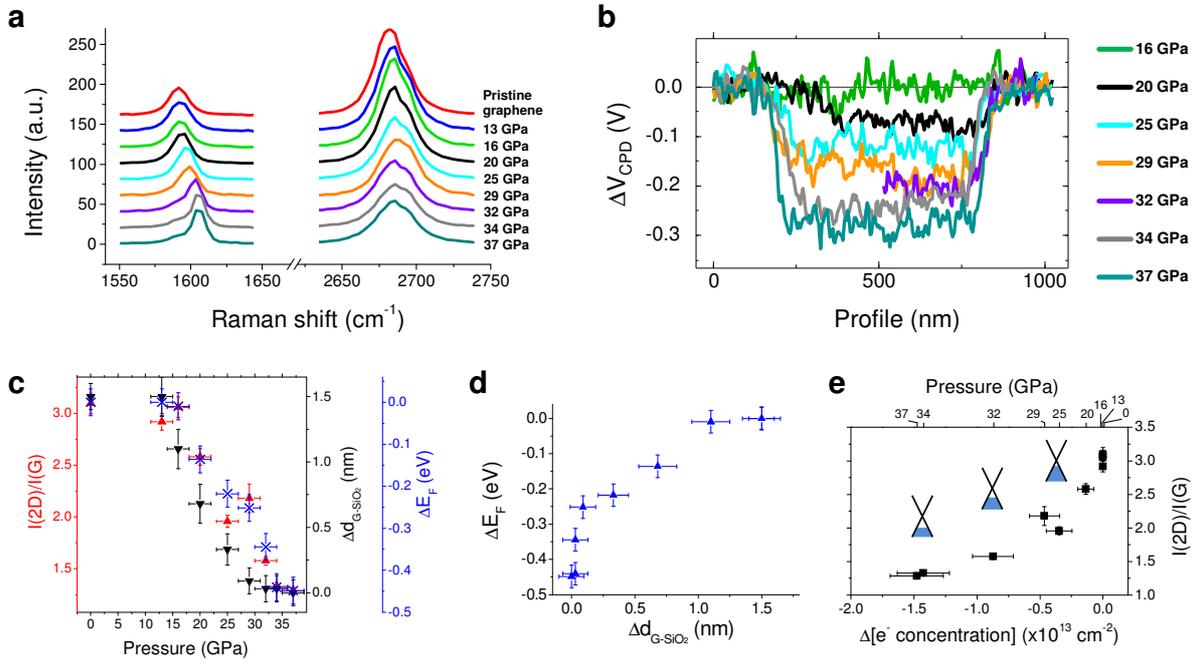

**Figure 2.** Evidence for local effective doping. a) G and 2D peaks of the Raman spectra from the different modified regions. b) Variation of the Contact Potential Difference (CPD) in the different regions. Curves obtained from selected profiles in Figure S4. c) 2D/G intensity ratio (red), graphene-SiO$_2$ distance variation (black) and Fermi level shift (blue) as functions of the pressure. d) Fermi level shift as a function of the graphene-SiO$_2$ distance variation. e) 2D/G intensity ratio as a function of the electron concentration. Insets show schematic sketches of the graphene band diagram representing the increase of hole doping with applied pressure. In c,d) $\Delta d_{G-SiO_2}$ accounts for the difference in pristine graphene and modified regions surface heights whereas $\Delta d_{G-SiO_2} = 0$ has been chosen as the minimum graphene-SiO$_2$ distance (corresponding to the deepest graphene modified area before breakage).

In addition to KPFM and Raman spectroscopy, we have further confirmed the change in the doping level by Scanning X-ray Photoelectron Microscopy (SPEM) measurements, comparing pristine and modified graphene regions (**Figure 3**). Figure 3a and 3b show AFM and C 1s SPEM images from a graphene flake with different modified areas. The SPEM images resolve the different regions, which present a clear contrast between pristine and modified areas at high pressures. Figure 3c shows a series of C 1s peaks corresponding to different applied pressures in the areas shown in Figure 3a and 3b, taken in the micro spot spectroscopic mode. The binding energy (284.75 eV) and shape of the pristine (0 GPa) graphene peak confirm the good quality of the graphene under study (see also Figure S7 and Table S1), when compared to values obtained for graphene on $SiO_2$ or SiC.[19] Figure 3c highlights that after applying ultrahigh pressure, the C 1s peaks present a rigid shift to lower binding energies, ranging from ~ 100 to 230 meV. This effect reflects a shift of the Fermi level towards the valence band, as expected in the case of a p-doping effect, in good agreement with Raman and KPFM observations. The rigid displacement of the C 1s peak is very robust, and always towards lower binding energies across all analyzed samples, confirming the p-doping effect. Note that the two spectra corresponding to the highest applied pressures do not follow the trend and are shifted to slightly smaller values than expected for a linear trend. Note also that the areas modified with the highest pressures (> 30 GPa) are seen in the SPEM image with a darker color, reflecting a lower C 1s intensity. The contrast is in part due to the shift of the C 1s peak, as the image is tuned to the maximum of the C 1s peak in the pristine graphene. However, an inspection of the C 1s core levels in Figure 3c shows that also the absolute intensity of the C 1s peaks decreases by ~ 10%, suggesting that the modified areas present a lower density of C atoms (see also Table S1 in the SI). This may indicate that graphene has an initial corrugation that is ironed out because of the ultrahigh pressure applied by the diamond tip.

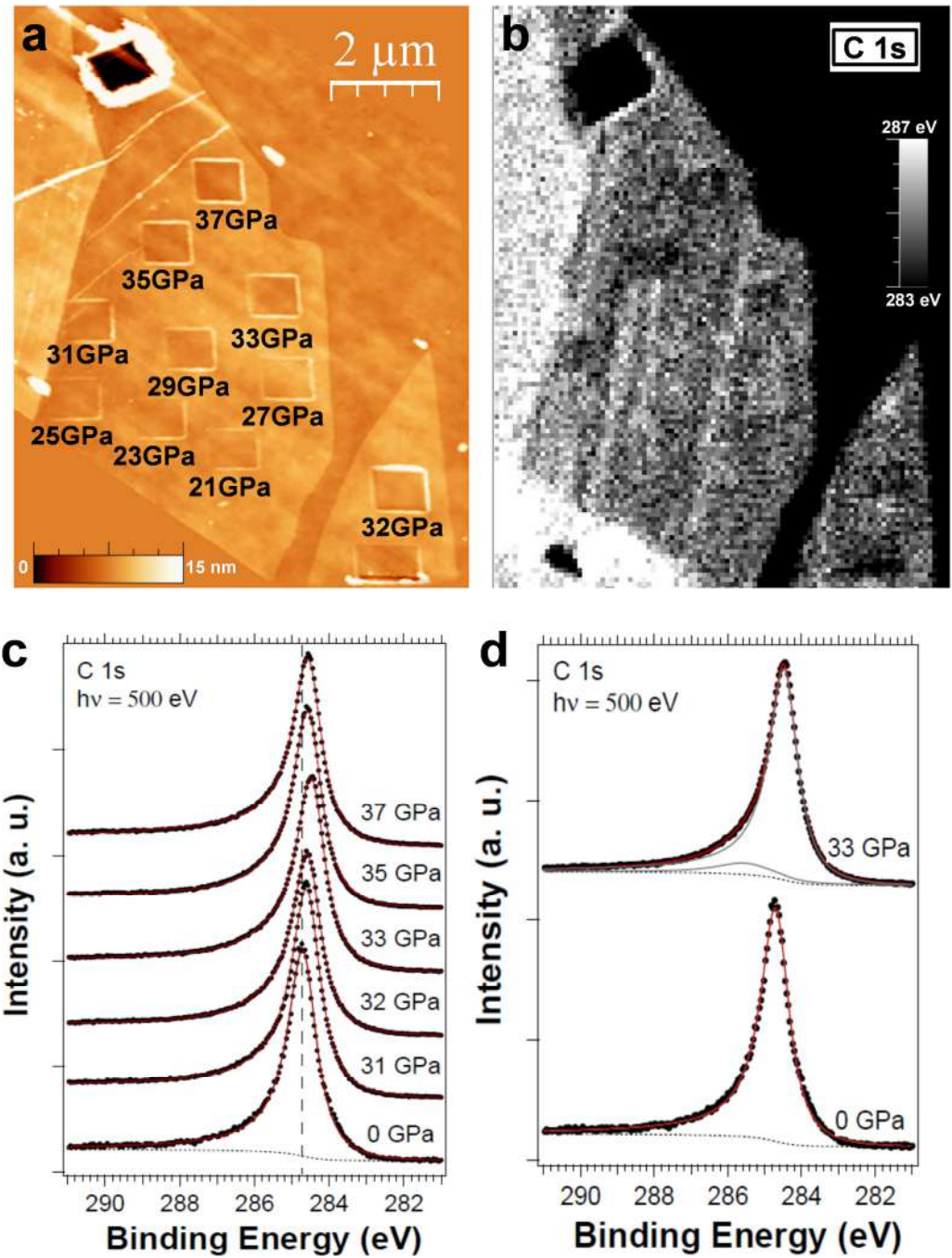

**Figure 3.** Scanning X-ray Photoelectron Microscopy. a) AFM topographic image showing 800 x 800 nm² areas modified under different pressures. b) C 1s SPEM image of the same area as in a) in greyscale (white means more intensity) taken with hν = 500 eV. c) and d) Representative C 1s peaks as a function of the applied pressure. The spectra are vertically shifted for the sake of clarity. Dots are experimental points, the red lines are fits to the data (see text). In c), the background is shown for the 0 GPa case. A vertical dashed line indicates the binding energy for 0 GPa, note the shift of the C 1s peak after pressure modification. In d) the deconvolution of the 0 GPa (one single component) and 33 GPa (two components) is shown. See also Figure S7 and S8.

## 3. Discussion

In order to explain the permanent changes in the height of the graphene flakes and the modification of the doping level observed in the experiments, we suggest that strong covalent bonds between graphene and the substrate are formed when ultrahigh pressure is applied. No doping effect is observed in few layer graphene areas (see Figure S9), presumably due to the weak interlayer coupling and the absence of bonds between carbon layers. Beyond van der Waals attraction, short-range interactions between graphene and defects in $SiO_2$ have indeed been shown as the origin of the naturally occurring ultrastrong adhesion between these two surfaces.[20] This idea of covalent bonding is inspired by the phase diagram transition from graphite to metastable diamond through amorphous carbon, which contains a growing number of $sp^3$ bonds as the pressure increases.[21] However, since graphene in the modified regions only shows a change in the doping level and, according to the Raman measurements, a minor presence of defects, we speculate that a small fraction of bonds form per unit area.

It is widely assumed that under ambient conditions there exists an ever-present layer of buffer molecules which, after graphene transfer on the substrate, remains captured between the flakes and $SiO_2$.[22] This buffer layer is responsible for most of the 1.3 nm apparent thickness of graphene in our AFM images. As demonstrated in ref. [22b], when visualizing graphene with forces of the order of 10 nN, the AFM tip presses the graphene through the entire buffer layer and onto the underlying substrate. Hence, in all our modifications, which involve forces in the µN range, we have brought graphene into hard contact with the $SiO_2$ substrate, completely expelling the underlying buffer molecules. This is evidence of the independence of our procedure from possible interfacial contaminants. However, interestingly, different behaviors are found when the applied load is removed. For the lowest pressures (< 16 GPa) the deformation is completely reversible, suggesting that the removed buffer layer can refill the volume beneath graphene, which recovers its "normal" height; in this regime, no strong covalent bonds are formed. It is important to remark that, as well as no change in height, there is no material accumulation on the sides of the modified areas, indicating that the effects observed with the ultrahigh pressure procedure are not related to the removal of possible surface contaminants. For pressure values in the 16-25 GPa range, the process is not completely reversed. For these pressure values, some covalent bonds might be formed, but in a very limited number. This fact, combined with the Gaussian roughness of the surface, allows a limited portion of molecules to return to the interstitial region. Finally, for higher pressures (> 25 GPa) the deformation becomes totally irreversible, implying that the buffer molecules are completely expelled from the modified areas and the graphene flake remains in contact with the $SiO_2$ substrate (Figure 1d). Parallel experiments with this technique allow better sealing of graphene blisters, reflected in a significant drop of the leak rates.[23] This confirms that graphene is brought into hard contact with the $SiO_2$ substrate and that its adhesion is significantly increased.

In order to verify our hypothesis, we carried out further studies of the graphene $SiO_2$ system. Firstly, we checked for the presence of chemically modified carbon atoms in our graphene sample, analyzing the line shape of the C 1s core levels shown in Figure 3 (see Experimental section and Figure S8). The line shape of the C 1s peak coming from pristine single-layer graphene presents a width (FWHM) ranging between 0.88 and 1.07 eV, depending on the specific flake probed (Figure S7). The asymmetry found in all cases is $\alpha = 0.14$. In the modified areas, a visual analysis of the C 1s line shape does not show any significant change (Figure 3c), besides the shift of the peak. As a first step, we tried to fit the C 1s line shape using a single component with the same parameters found for pristine graphene. This gives rise to the largest component shown in Figure 3d for 33 GPa. However, a second smaller component is needed to explain the line shape broadening on the left of the main component. As the new component is very small, its fitting parameters (binding energy, width, etc.) cannot be determined with high accuracy. The fit is made leaving the binding energy and the intensity as free parameters, and using the same values for the width and asymmetry as for the main component. This procedure results in a second component shifted by 1.03 eV towards higher binding energy. All C 1s core levels from modified regions present a second component (Figure 3d and S8). Its intensity grows with the applied pressure, ranging from 5 to 10% of the total intensity, with the exception of the two spectra corresponding to the highest pressures applied (35 and 37 GPa), which also present a smaller binding energy shift of the main component of the C 1s core level. The values reported for C atoms in sp$^3$ hybridization induced in graphene range from 0.8 eV[24] to 1.0 eV[19c] (in both cases possibly bound to H), to be compared with a standard core level shift of 0.8-0.9 eV between C 1s in graphite and diamond.[25] C 1s from graphene bound to other species has much larger core level shifts: 2.1 eV (C-O-C), 3.7 eV (C=O), 5.1 eV (O-C=O),[26] or 1.7 eV for C-OH.[27] Since the binding energy shift of the second component was 1.03 eV, we conclude that the second component observed is compatible with carbon atoms in sp$^3$ hybridization.

Theoretical simulations support the hypothesis suggested by the experimental results. Looking at the graphene SiO$_2$ coupling mechanism reported in the existing literature, we face a puzzling scenario. According to some authors,[28] the interaction between graphene and SiO$_2$ should be weak and exclusively induced by van der Waals forces, with graphene stabilized at around 3 Å from the SiO$_2$ surface without any appreciable charge transfer (physisorption). Other authors[29] conclude that strong covalent bonds between the C atoms and the substrate atoms are formed (chemisorption), and, as a consequence, the Dirac cone structure is lost. Finally, both covalent bonding and physisorption of graphene on SiO$_2$ are suggested in other works.[30] To shed some additional light on the binding mechanism, we carried out DFT-based simulations of adsorption of graphene on SiO$_2$ with two main goals: firstly, to clarify whether or not it is possible to induce the formation of covalent bonds between the graphene atoms and the SiO$_2$ atoms by applying ultrahigh pressure; and secondly, to estimate a minimum number of covalent bonds needed to maintain the graphene flake strongly coupled to the surface.

Instead of a classical DFT adsorption study, we run hundreds of geometry optimizations, using *extreme* starting positions, looking for local minima in the configuration space in which some of the carbon atoms make covalent bonds with the SiO$_2$ substrate. In this way we found four different chemisorbed configurations of graphene (see Figure S14 and the corresponding discussion in the SI). Starting from them, we constructed continuous paths, in the configuration space, between the covalently bonded geometries and the geometry of the "perfect" graphene sheet placed far away from the $SiO_2$ surface. Then, we performed single point DFT calculations along these continuous paths, estimating the corresponding pressure, as described in the Experimental section. As shown in **Figure 4a**, the pressure profiles present a barrier of the order of 10-20 GPa, in excellent agreement with the pressure threshold observed in the experiments. Thus, theoretical simulations confirm that the pressure used in the experiments is high enough to induce the formation of chemical bonds between the graphene layer and the $SiO_2$ substrate.

In order to assess whether irreversible chemical contact between graphene and the $SiO_2$ substrate is possible when only a small fraction of C atoms are involved in the bonds, we performed further geometry optimizations using (2×2), (3×3), and (4×4) $SiO_2$ supercells. In the starting position, the graphene layer was chemically bonded to the substrate only in a small portion of the unit cell, while the remaining dangling bonds were saturated by H atoms. In this way, we have been able to test the strength of the C-surface bonds, because the graphene layer is on average repelled by the surface in all the regions where no chemical bonds are present (see SI for a detailed discussion). The chemical bonds were preserved for the "Si-O-O-1N" and the "Superdense" configurations. In Figure 4b, we show the side views of the final coordinates of the supercell geometry optimization in the "Si-O-O-1N". We carry out a more quantitative analysis in Figure 4c, in which the C atoms are grouped according to their distance from the surface ($z-z_s$). In all the supercells, we identify a group of two C atoms chemically bonded to the O atoms of the substrate. Starting from the (2×2) supercell, we find two groups composed of 4 and 10 C atoms, which remain at roughly the same distance from the surface. Finally, we have a group that contains all the remaining C atoms (the "faraway atoms"), whose average distance increases as the size of the supercell increases and tends to the equilibrium distance of graphene stabilized by van der Waals forces. From the (4×4) supercell results, we can directly estimate that ~ 1% of chemically bonded C atoms are sufficient to keep the graphene layer covalently bonded to the $SiO_2$ substrate, the $sp^2/sp^3$ ratio being of the order of 5%. Furthermore, we can infer that the provided figures are upper bounds, because in the (4×4) supercell the faraway atoms almost reached the physisorption configuration, indicating that constructing a bigger supercell would not induce more stress on the C atoms participating in the chemical bonds. This conclusion agrees fairly well with the experimental results, indicating that the amount of bonds created is indeed very small. Finally, from the DFT calculations, we estimated the surface graphene charge transfer, finding that for a low density of covalent bonds, the far away areas present a p-doping compatible with a Fermi level shift of the order of -0.2 eV.

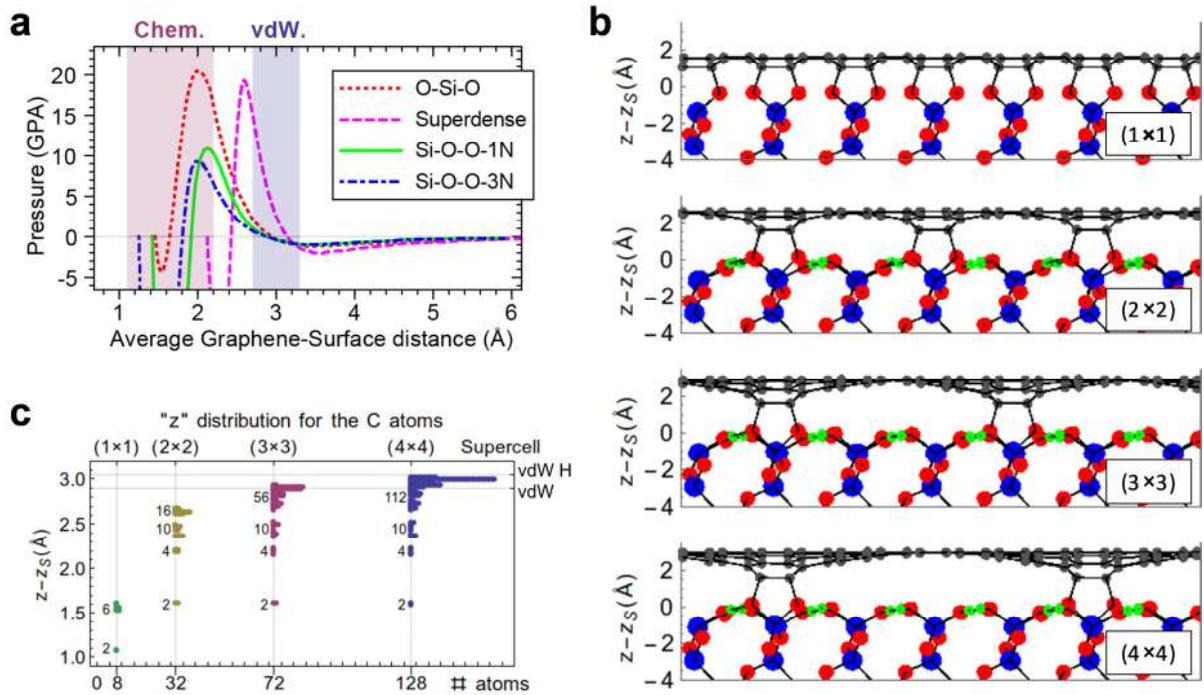

**Figure 4.** DFT calculations. a) Pressure barrier to reach chemisorption as obtained for the four chemisorbed configurations analyzed in this work. The shaded region labeled Chem. (vdW.) indicates the graphene-surface distance range in which chemical bonding (van der Waals attraction) is usually observed. b) Side view after the geometry optimization for the supercells discussed in the main text. c) z value distribution of the C atoms after the geometry optimization; the horizontal grid-lines mark the average z position of the C atoms in the van der Waals attraction configuration without the H atoms (vdW) and for the fully saturated surface (vdW H); the numbers inside the plot indicate the total number of C atoms in the corresponding group of histograms.

The analysis performed is an idealization of a quite complex scenario, characteristic of experiments in air ambient conditions, where the number of parameters is much higher than in, for instance, ultrahigh vacuum experiments. Nevertheless, it presents the benefit of closer conditions with respect to technological applications. First of all, our amorphous $SiO_2$, a substrate with technological relevance, presents a statistical roughness higher than 0.5 nm. Secondly, the modified areas involve millions of atoms, with the probable presence of oxygen and hydrogen terminations or hydroxyl groups.[28b] In this respect, it is worth mentioning that several other effects could induce p-doping. Charge accumulation, originating from possible gating of the $SiO_2$ on the graphene by modification of the electronic state of $SiO_2$, might be occurring. Experiments on the evolution of charges in graphene/$SiO_2$ systems in ambient conditions show that the decay of the injected charges through the water adsorbed on the $SiO_2$ surface occurs in minutes, a few hours in the extreme cases.[31] For electrochemical strain effects, the characteristic diffusion times are of the order of ~1 s.[32] Since our KPFM and Raman spectroscopy data were acquired two months after sample modifications (see Experimental section), in our case, these effects should be very weak. Another important effect to consider is flexoelectricity, which in general terms is the response of polarization to a strain gradient.[33] These strain gradients are known to be huge under an AFM tip, especially in our modifications, where we are applying extremely high loads. However, flexoelectricity is proportional to the permittivity of the material.[33] We observe that $SiO_2$ and graphene permittivities are low ($\varepsilon_{SiO2}$ ~ 4, $\varepsilon_{Gr}$ ~ 2-15),[34] whereas the permittivities of materials such as $PbTiO_3$ or $BaTiO_3$, where flexoelectric effects have been observed,[33] are several orders of magnitude higher ($\varepsilon$ ~$10^2$-$10^4$). Thus, although flexoelectric effects might be taking place, their effects on the effective doping should be weak. Indeed, the KPFM signal of $SiO_2$ acquired over regions modified under ultrahigh pressures (Figure S10) shows a negligible change as compared to the one obtained in pristine zones, thus confirming that its electronic configuration is not modified by the strain induced by the tip.

We can envision some practical applications. In addition to the above-mentioned improved sealing of graphene blisters,[23] a direct consequence of the doping in graphene is the creation of low electrical contact resistance areas,[35] already observed in conventional electrostatically-doped graphene in transistor configurations.[14] In order to confirm that the ultrahigh pressure modification of the graphene doping level was reflected in the electrical contact resistance, we carried out conductive AFM (C-AFM) experiments on our samples (**Figure 5**). In these measurements, a first macroscopic electrode was in electrical contact with the graphene flake (see inset in Figure 5b). A metal-coated AFM tip acted as a second mobile electrode, which allowed mapping of the current variations in the different regions of the graphene flake (see Experimental section for further details). Figure 5a and 5b show simultaneous topography and current maps of modified areas under a 35 GPa pressure, indicating a clear increase in the measured current. Since the current is simultaneously acquired in both pristine and modified regions, with exactly the same scanning parameters and bias voltage, we can attribute this current increase to a decrease of the contact resistance between the tip and the modified areas. From the current variation in Figure 5d, a sizable ~35% improvement for the resistance is found.

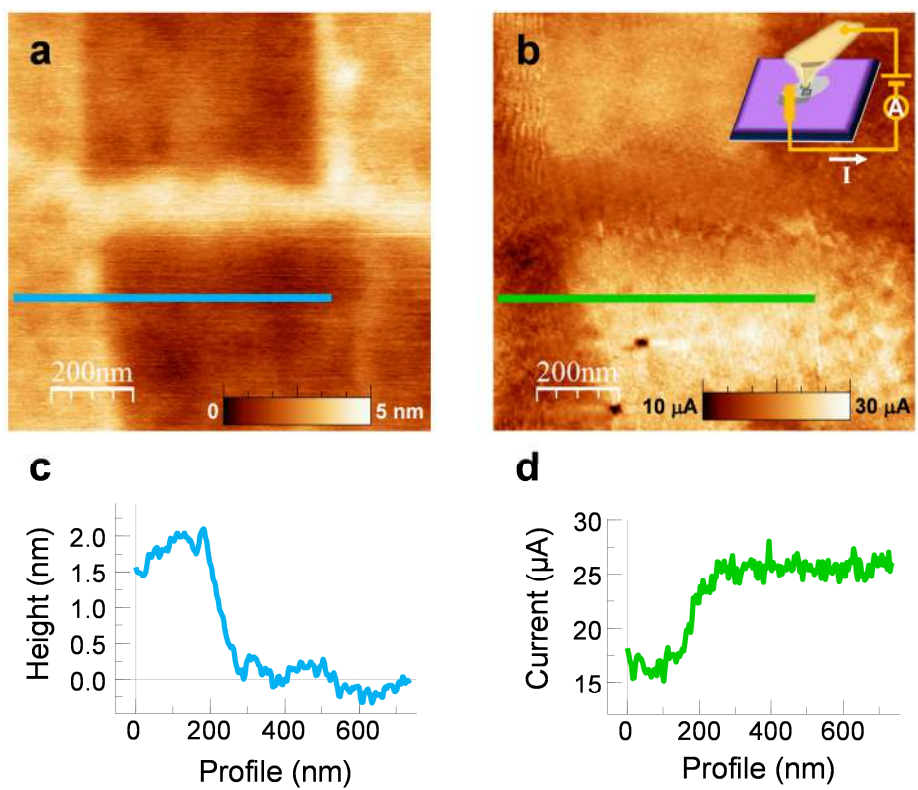

**Figure 5.** Conductive-AFM on 35 GPa modified areas. a) Topographic image. b) Current map at a fixed Bias voltage of 0.6 V. c) and d) Profiles along the lines in a) and b) respectively. The inset in b) is a schematic representation of the electrical circuit.

## 4. Conclusion

To end up, we would like to remark that the presented method could be extended to other 2D materials and van der Waals heterostructures and related systems. The improvement of the contact resistance in selected areas through the controlled doping effect opens the possibility of obtaining similar currents but using lower voltages, which might allow a remarkable reduction of the power consumption of future graphene-based electronic devices. Additionally, we carried out parallel experiments, which are referred in the manuscript, that show that this technique allows better sealing of graphene blisters, reflected in a significant drop of their leak rates. This result would represent a direct improvement in devices such as membrane sensors. Hence, our results might have a direct impact in different fields, such as better flexible electrodes, batteries or sensors, to name but a few. Finally, thanks to the remarkable mechanical properties of graphene, our method opens an avenue to carry out chemical reactions of trapped molecules between graphene layers and a substrate at ultrahigh pressures in a simple and quick manner, without the technical drawbacks characteristic of classical high pressure procedures. We are exploring further technological applications for this technique.

**5. Experimental Section**

*Pressure calculation.* A simple estimation of the pressure was performed following the Hertz model.[36] Figure S11 shows schematics of the AFM setup with a diamond tip. The contact radius $\rho$ between a sphere and a plane is then given by:

$$\rho = \left(\frac{3FR}{4E^*}\right)^{1/3} \qquad (3)$$

where $F$ is the applied force, $R$ the tip radius and $E^*$ is the effective elastic modulus $1/E^* = 1/E_{tip} + 1/E_{sample}$. The mean contact pressure $P$ exerted on the sample by the tip was determined as:

$$P = \frac{1}{\pi}\left(\frac{4E^*}{3}\right)^{2/3}\left(\frac{F}{R^2}\right)^{1/3} \qquad (4)$$

Using single crystal diamond AFM probes, with cantilever stiffness $k \sim 40$ N m$^{-1}$, a conventional AFM set up can easily apply forces higher than 100 µN. Assuming $E^*$ of the order of 1 TPa and a tip radius of ~ 50 nm, the resulting pressure is well above 100 GPa. By using sharper tips and stiffer cantilevers (as for instance commercially available $R \sim 20$ nm and $k \sim 80$ N m$^{-1}$ (www.nanosensors.com)) this figure could reach values up to ~ 300 GPa. We employed a single crystal diamond tetrahedral pyramid SCD15/AIBS probe from MikroMasch (www.spmtips.com). The nominal cantilever length and width were 125 and 35 µm respectively. It presented a resonance frequency of 320 kHz in air, a Q factor of 480 and a force constant of 32 N m$^{-1}$ (calibrated using the Sader method[37]). The nominal tip radius of the SCD15/AIBS probes was < 10 nm, although this value was not reliable since the tip apex was easily splintered, reflected in unstable images. We performed force *vs.* distance curves and scans in contact mode at high loads on a diamond surface until the tip presented stable imaging. At this point, we calibrated the tip radius by imaging carbon nanotubes of different heights,[38] obtaining a value of $R$ = 42 nm (Figure S12) that remained stable for months. We have also successfully tried high-density diamond like carbon spherical tips B50 from Nanotools (www.nanotools.com), with a nominal force constant of 40 N m$^{-1}$ and a nominal tip radius of 50 (± 5) nm, obtaining similar results (see Figure S13), although their long-term durability compared to the single crystal diamond tips is lower.

*Atomic Force Microscopy (AFM).* We carried out AFM measurements using a Cervantes Fullmode AFM from Nanotec Electronica SL. We employed WSxM software (www.wsxm.es) both for data acquisition and image processing.[39] In order to avoid sample damage, we took topographic images before and after modifications in dynamic mode with an amplitude set point of 25 nm (cantilever free amplitude 30 nm).

We determined the deflection sensitivity of the optical detection system by performing force *vs.* distance curves on the substrate, which we assumed was non-deformable in the distance range we used, and measuring the slope of the curves in the contact regime.[40] We typically obtained sensitivity values of ~ 30 nm/V. We calibrated the AFM scanner with two gratings with pitch distances 3 μm (TGX01, Mikromasch) and 300 nm (2D300, Nanosensors), and Highly Oriented Pyrolytic Graphite (HOPG, SPI supplies), with an atomic periodicity of 0.25 nm. The X-Y calibration parameters of the piezo scanner were found to depend linearly on the scanning size, ranging from 29 nm/V for sub-micrometer scan sizes to 36 nm/V for the scanner maximum scan size (~ 12 μm). The result of the linear fit was introduced in the control software of the AFM, which automatically adapted the piezo driving voltages to the required scanning size. The Z direction was calibrated using HOPG's atomic step height (0.34 nm), finding a calibration of 12 nm/V.

*Raman spectroscopy.* We acquired Raman spectra two months after sample modifications using a WITEC/ALPHA 300RA Raman confocal microscope (Witec GmbH, Ulm, Germany) at ambient conditions. The setup was equipped with a 600 lines mm$^{-1}$ grating and a Nikon 100× objective (NA = 0.95), which enabled micro Raman with a laser spot diameter of ~ 300 nm. Samples were mounted on a piezoelectric stage, providing nanometer resolution to acquire spectra in very precise locations. The laser wavelength and power were 532 nm and 1 mW respectively. We used an integration time of 0.5 s per spectrum and a single accumulation for all the measurements. Raman mappings were carried out under these conditions with a step size of 100 nm. For the spectra shown in Figure 2a, we acquired Raman mappings and averaged the spectra from the central areas (~ 300 x 300 nm$^2$) of the modified regions, which meant averaging 9 spectra.

*Kelvin Probe Force Microscopy (KPFM).* The same setup used for AFM imaging allowed KPFM measurements. We employed ElectriMulti75-G probes from BudgetSensors (www.budgetsensors.com), with nominal stiffness and resonance frequency of 3 N m$^{-1}$ and 75 kHz respectively. We performed simultaneous dynamic mode AFM for the topography and Frequency Modulation mode for the KPFM.[41] For the topography, we used an amplitude set point of 15 nm (cantilever free amplitude 20 nm), whereas for the CPD we employed an alternate bias voltage of amplitude 3.5 V and frequency 7 kHz. To have a well-defined potential reference, we electrically grounded the flake through the Exfoliated Graphite Flakes (EGF) soft-electrode procedure[42] to provide a stable potential reference. We performed KPFM measurements two months after sample modifications in an inert Ar atmosphere to avoid CPD shielding by the presence of an adsorbed water layer on the surface.[43]

*Density Functional Theory (DFT) calculations.* We performed DFT calculations using the projector augmented wave (PAW) method,[44] as implemented in the plane-wave based code VASP.[45] We adopted the generalized gradient approximation (GGA), using the PBE functional,[46] to describe the exchange-correlation energy. We used the PAW method to represent the core electrons and a high plane wave cut-off (700 eV) to account for short C-O bonds. We included weak dispersion forces (necessary to properly describe graphene stabilized by van der Waals attraction) using the Tkatchenko-Sheffler method.[47] In all calculations, we set the tolerance in the electronic self-consistent cycle to 10$^{-5}$ eV, using Gamma centered Monkhrost Pack grids[48] characterized by (at least) $\Delta k < 0.3$ Å$^{-1}$. We modeled the SiO$_2$ surface by constructing (001)-oriented slabs of α-quartz (lattice constants a = 4.916 Å, c = 5.405 Å), which offered the advantage of having hexagonal symmetry. Model slabs included 6 SiO$_2$ units. We checked the robustness of the results in control calculations using 12 SiO$_2$ units, as well as in calculations with different surface terminations, including H atoms to saturate the dangling bonds, or including defects. The best lattice constant matching between graphene and SiO$_2$ was achieved using a (1×1) in plane SiO$_2$ cell and a (2×2) graphene supercell, containing 8 C atoms. To cope with the fact that we carried out the experiments using an amorphous surface, we placed graphene over 4 different surface terminations (see Figure S14 and the corresponding discussion in the SI), characterized by being either highly reactive, or fairly stable.[49] For the adsorption calculations, we chose not to saturate the dangling bonds (where applicable) with H atoms in the surface on which the graphene layer was to be adsorbed. In the supercell calculations, we chose the starting positions as follows: in a small fraction of the supercell (a (1×1) SiO$_2$ unit cell) the graphene coordinates were set to the position of the covalent bonds. In the rest of the cell, the dangling bonds of the substrate were saturated by H atoms, whereas the coordinates of the graphene atoms were gradually set to higher distances from the surface. In all the calculations used in this work, we have determined the final atomic arrangement by a full geometry relaxation of all the coordinates of graphene and of the topmost SiO$_2$ unit in the slab, which means (depending on the surface terminations) at least 1 Si layer, 2 O layers and all H atoms, where applicable. We continued the geometry relaxations up to the point at which the maximum total force on the active atoms was less than 0.01 eV/Å.

*Estimation of pressure through DFT.* Given a geometry configuration in which graphene (perfect or distorted) is placed over SiO$_2$, the pressure necessary to keep graphene in that configuration can be estimated. From the DFT calculation, one can get the forces acting on each C atom. Summing the *z*-component (*i.e.* the out-of-plane component) of the forces over all the C atoms and dividing the result by the area of the in-plane unit cell, we obtain a quantity with the dimension of a pressure. It is worth mentioning that, by definition, such pressure estimations might give (non-physical) negative values. These describe the situation in which graphene is attracted by the surface, as in the vicinity of the covalently bonded or van der Waals attraction equilibrium configurations, where, by definition, we have p = 0. Finally, in all pressure *vs.* z plots, we determined the z coordinate as the z average over all C atoms.

*Scanning X-ray Photoelectron Microscopy (SPEM).* We performed the experiments at the ESCAmicroscopy beamline,[50] receiving synchrotron light from the Elettra storage ring in Trieste (Italy). The monochromatic X-ray beam was focused on the sample using a Fresnel zone plate combined with an order-selecting aperture. The sample-illuminated area had a diameter around 150 nm and was raster scanned with respect to the microprobe. Photoelectrons were collected with a Specs-Phoibos 100 hemispherical analyzer and detected using a 48-channel electron detector.[51] The electron analyzer axis was fixed at 30º with respect to the surface normal, in order to enhance the surface sensitivity of the instrument.

SPEM can be performed in imaging spectromicroscopy and micro spot spectroscopic modes. The imaging spectromicroscopy mode probes the lateral distribution of elements by collecting photoelectrons with a selected kinetic energy window, while scanning the specimen with respect to the microprobe. The spatial variation in the contrast of the images reflects the variation of the photoelectron yield, which is a measure of the local concentration of the corresponding chemical species. The micro spot mode is identical to the conventional X-ray Photoelectron Spectroscopy (XPS) technique. In this mode, energy distribution curves were measured from the illuminated local micro spot area. In general, spectra in the micro spot mode had longer acquisition times and a better signal-to-noise ratio than SPEM images. In this work, we acquired spatially resolved photoemission spectra of selected regions and chemical maps with 0.2 eV energy resolution by using 500 eV photon energy. The core level binding energies were calibrated using an Au reference.

*Deconvolution of C 1s Core Levels*: We fitted the line shape of C 1s core levels using a Shirley background and asymmetric singlet pseudo-Voigt functions. We optimized the fit using a Levenberg-Marquardt algorithm with a routine running in IGOR Pro (WaveMatrix Inc.).[52] We judged the quality of the fit from a reliability factor, the normalized $\chi^2$.

*Conductive AFM (C-AFM).* The same setup used for AFM imaging and KPFM measurements was equipped with a homemade IV preamplifier allowing C-AFM measurements. As for KPFM, we employed ElectriMulti75-G probes. We obtained current maps by scanning in Contact mode with a Normal force set point of 150 nN and a fixed Bias voltage of 0.6 V. Note that the electrical resistance measured with this technique is always an upper bound for the contact resistance since potential contamination of the tip/surface reduces tip-sample current.

**Acknowledgements**
We thank A. del Campo and the ESCAmicroscopy beamline team at Elettra for technical assistance with Raman and SPEM measurements respectively. We also thank A. Gil for insightful discussions and H. Nevison-Andrews for a careful reading of the manuscript. We acknowledge computer time provided by the Centro de Computación Científica of the Universidad Autónoma de Madrid and the Red Española de Supercomputación. We acknowledge financial support from the Spanish Ministry of Economy and Competitiveness, through the "María de Maeztu" Programme for Units of Excellence in R&D (MDM-2014-0377), the projects MAT2016-77608-C3-1-P and -C3-3-P, MAT2014-52477-C5-5-P, FIS2016-77889-R and the Comunidad de Madrid through the programme Nanofrontmag and MAD2D-CM Program. C. Díaz acknowledges the Ramón y Cajal program of the MINECO.

**References**

[1]     A. Jayaraman, *Rev. Mod. Phys.* **1983**, *55*, 65.
[2]     L. Dubrovinsky, N. Dubrovinskaia, E. Bykova, M. Bykov, V. Prakapenka, C. Prescher, K. Glazyrin, H. P. Liermann, M. Hanfland, M. Ekholm, Q. Feng, L. V. Pourovskii, M. I. Katsnelson, J. M. Wills, I. A. Abrikosov, *Nature* **2015**, *525*, 226.
[3]     C. Lee, X. D. Wei, J. W. Kysar, J. Hone, *Science* **2008**, *321*, 385.
[4]     K. S. Novoselov, V. I. Fal'ko, L. Colombo, P. R. Gellert, M. G. Schwab, K. Kim, *Nature* **2012**, *490*, 192.
[5]     a) C. H. Y. X. Lim, M. Nesladek, K. P. Loh, *Angew. Chem. Int. Ed.* **2014**, *53*, 215; b) K. S. Vasu, E. Prestat, J. Abraham, J. Dix, R. J. Kashtiban, J. Beheshtian, J. Sloan, P. Carbone, M. Neek-Amal, S. J. Haigh, A. K. Geim, R. R. Nair, *Nat. Commun.* **2016**, *7*, 12168.
[6]     a) N. Domingo, L. Lopez-Mir, M. Paradinas, V. Holy, J. Zelezny, D. Yi, S. J. Suresha, J. Liu, C. R. Serrao, R. Ramesh, C. Ocal, X. Marti, G. Catalan, *Nanoscale* **2015**, *7*, 3453; b) K. Sotthewes, P. Bampoulis, H. J. W. Zandvliet, D. Lohse, B. Poelsema, *ACS Nano* **2017**, *11*, 12723.
[7]     a) J. Nicolle, D. Machon, P. Poncharal, O. Pierre-Louis, A. San-Miguel, *Nano Lett.* **2011**, *11*, 3564; b) S. Ryu, L. Liu, S. Berciaud, Y. J. Yu, H. T. Liu, P. Kim, G. W. Flynn, L. E. Brus, *Nano Lett.* **2010**, *10*, 4944; c) M. Yankowitz, J. Jung, E. Laksono, N. Leconte, B. L. Chittari, K. Watanabe, T. Taniguchi, S. Adam, D. Graf, C. R. Dean, *Nature* **2018**, *557*, 404.
[8]     M. Yankowitz, S. Chen, H. Polshyn, K. Watanabe, T. Taniguchi, D. Graf, A. F. Young, C. R. Dean, *arXiv:1808.07865* **2018**.

[9] A. C. Ferrari, J. C. Meyer, V. Scardaci, C. Casiraghi, M. Lazzeri, F. Mauri, S. Piscanec, D. Jiang, K. S. Novoselov, S. Roth, A. K. Geim, *Phys. Rev. Lett.* **2006**, *97*, 187401.
[10] a) A. Klemenz, L. Pastewka, S. G. Balakrishna, A. Caron, R. Bennewitz, M. Moseler, *Nano Lett.* **2014**, *14*, 7145; b) B. Vasic, A. Matkovic, U. Ralevic, M. Belic, R. Gajic, *Carbon* **2017**, *120*, 137.
[11] L. G. Cancado, A. Jorio, E. H. Martins Ferreira, F. Stavale, C. A. Achete, R. B. Capaz, M. V. O. Moutinho, A. Lombardo, T. S. Kulmala, A. C. Ferrari, *Nano Lett.* **2011**, *11*, 3190.
[12] A. Das, S. Pisana, B. Chakraborty, S. Piscanec, S. K. Saha, U. V. Waghmare, K. S. Novoselov, H. R. Krishnamurthy, A. K. Geim, A. C. Ferrari, A. K. Sood, *Nat. Nanotechnol.* **2008**, *3*, 210.
[13] S. Pisana, M. Lazzeri, C. Casiraghi, K. S. Novoselov, A. K. Geim, A. C. Ferrari, F. Mauri, *Nat. Mater.* **2007**, *6*, 198.
[14] Y. J. Yu, Y. Zhao, S. Ryu, L. E. Brus, K. S. Kim, P. Kim, *Nano Lett.* **2009**, *9*, 3430.
[15] W. Melitz, J. Shen, A. C. Kummel, S. Lee, *Surf. Sci. Rep.* **2011**, *66*, 1.
[16] R. Wang, S. N. Wang, D. D. Zhang, Z. J. Li, Y. Fang, X. H. Qiu, *ACS Nano* **2011**, *5*, 408.
[17] Y. J. Kim, Y. Kim, K. Novoselov, B. H. Hong, *2D Mater.* **2015**, *2*, 042001.
[18] H. Y. Yuan, S. Chang, I. Bargatin, N. C. Wang, D. C. Riley, H. T. Wang, J. W. Schwede, J. Provine, E. Pop, Z. X. Shen, P. A. Pianetta, N. A. Melosh, R. T. Howe, *Nano Lett.* **2015**, *15*, 6475.
[19] a) J. D. Emery, B. Detlefs, H. J. Karmel, L. O. Nyakiti, D. K. Gaskill, M. C. Hersam, J. Zegenhagen, M. J. Bedzyk, *Phys. Rev. Lett.* **2013**, *111*, 215501; b) D. Ferrah, J. Penuelas, C. Bottela, G. Grenet, A. Ouerghi, *Surf. Sci.* **2013**, *615*, 47; c) R. Zhang, Z. S. Wang, Z. D. Zhang, Z. G. Dai, L. L. Wang, H. Li, L. Zhou, Y. X. Shang, J. He, D. J. Fu, J. R. Liu, *Appl. Phys. Lett.* **2013**, *102*, 193102.
[20] S. Kumar, D. Parks, K. Kamrin, *ACS Nano* **2016**, *10*, 6552.
[21] Y. Gao, T. F. Cao, F. Cellini, C. Berger, W. A. de Heer, E. Tosatti, E. Riedo, A. Bongiorno, *Nat. Nanotechnol.* **2018**, *13*, 133.
[22] a) K. S. Novoselov, D. Jiang, F. Schedin, T. J. Booth, V. V. Khotkevich, S. V. Morozov, A. K. Geim, *Proc. Natl. Acad. Sci. U. S. A.* **2005**, *102*, 10451; b) C. J. Shearer, A. D. Slattery, A. J. Stapleton, J. G. Shapter, C. T. Gibson, *Nanotechnology* **2016**, *27*, 125704.
[23] Y. Manzanares-Negro, P. Ares, M. Jaafar, G. Lopez-Polin, C. Gomez-Navarro, J. Gomez-Herrero, *arXiv:1809.03786* **2018**.
[24] S. Rajasekaran, F. Abild-Pedersen, H. Ogasawara, A. Nilsson, S. Kaya, *Phys. Rev. Lett.* **2013**, *111*, 085503.
[25] a) J. Diaz, G. Paolicelli, S. Ferrer, F. Comin, *Phys. Rev. B* **1996**, *54*, 8064; b) Y. Mizokawa, T. Miyasato, S. Nakamura, K. M. Geib, C. W. Wilmsen, *J. Vac. Sci. Technol., A* **1987**, *5*, 2809.
[26] J. T. Chen, G. A. Zhang, B. M. Luo, D. F. Sun, X. B. Yan, Q. J. Xue, *Carbon* **2011**, *49*, 3141.
[27] N. Peltekis, S. Kumar, N. McEvoy, K. Lee, A. Weidlich, G. S. Duesberg, *Carbon* **2012**, *50*, 395.
[28] a) Z. Ao, M. Jiang, Z. Wen, S. Li, *Nanoscale Res. Lett.* **2012**, *7*, 1; b) W. Gao, P. Xiao, G. Henkelman, K. M. Liechti, R. Huang, *J. Phys. D: Appl. Phys.* **2014**, *47*, 255301; c) R. H. Miwa, T. M. Schmidt, W. L. Scopel, A. Fazzio, *Appl. Phys. Lett.* **2011**, *99*, 163108; d) T. C. Nguyen, M. Otani, S. Okada, *Phys. Rev. Lett.* **2011**, *106*, 106801.
[29] a) J. Dai, J. Yuan, *Chem. Phys.* **2012**, *405*, 161; b) Y.-J. Kang, J. Kang, K. J. Chang, *Phys. Rev. B* **2008**, *78*, 115404; c) P. Shemella, S. K. Nayak, *Appl. Phys. Lett.* **2009**, *94*, 032101
[30] a) R. Colle, G. Menichetti, G. Grosso, *Phys. Status Solidi B* **2016**, *253*, 1799; b) X. F. Fan, W. T. Zheng, V. Chihaia, Z. X. Shen, J.-L. Kuo, *J. Phys.: Condens. Matter* **2012**, *24*, 305004.
[31] a) M. Konecny, M. Bartosik, J. Mach, V. Svarc, D. Nezval, J. Piastek, P. Prochazka, A. Cahlik, T. Sikola, *ACS Appl. Mater. Interfaces* **2018**, *10*, 11987; b) A. Verdaguer, M.

Cardellach, J. J. Segura, G. M. Sacha, J. Moser, M. Zdrojek, A. Bachtold, J. Fraxedas, *Appl. Phys. Lett.* **2009**, *94*, 233105
[32]  S. V. Kalinin, A. N. Morozovska, *J. Electroceram.* **2014**, *32*, 51.
[33]  G. Catalan, A. Lubk, A. H. G. Vlooswijk, E. Snoeck, C. Magen, A. Janssens, G. Rispens, G. Rijnders, D. H. A. Blank, B. Noheda, *Nature Mater.* **2011**, *10*, 963.
[34]  E. J. G. Santos, E. Kaxiras, *Nano Lett.* **2013**, *13*, 898.
[35]  a) J. T. Smith, A. D. Franklin, D. B. Farmer, C. D. Dimitrakopoulos, *ACS Nano* **2013**, *7*, 3661; b) F. N. Xia, V. Perebeinos, Y. M. Lin, Y. Q. Wu, P. Avouris, *Nat. Nanotechnol.* **2011**, *6*, 179.
[36]  K. L. Johnson, *Contact Mechanics*, Cambridge University Press, Cambridge **1985**.
[37]  a) J. E. Sader, R. Borgani, C. T. Gibson, D. B. Haviland, M. J. Higgins, J. I. Kilpatrick, J. N. Lu, P. Mulvaney, C. J. Shearer, A. D. Slattery, P. A. Thoren, J. Tran, H. Y. Zhang, H. R. Zhang, T. Zheng, *Rev. Sci. Instrum.* **2016**, *87*, 093711; b) J. E. Sader, J. W. M. Chon, P. Mulvaney, *Rev. Sci. Instrum.* **1999**, *70*, 3967.
[38]  a) P. Markiewicz, M. C. Goh, *Langmuir* **1994**, *10*, 5; b) A. D. Slattery, C. J. Shearer, J. G. Shapter, J. S. Quinton, C. T. Gibson, *Nanomaterials* **2017**, *7*, 346; c) Y. Wang, X. Y. Chen, *Ultramicroscopy* **2007**, *107*, 293.
[39]  I. Horcas, R. Fernandez, J. M. Gomez-Rodriguez, J. Colchero, J. Gomez-Herrero, A. M. Baro, *Rev. Sci. Instrum.* **2007**, *78*, 013705.
[40]  A. D. Slattery, A. J. Blanch, J. S. Quinton, C. T. Gibson, *Ultramicroscopy* **2013**, *131*, 46.
[41]  T. Glatzel, S. Sadewasser, M. C. Lux-Steiner, *Appl. Surf. Sci.* **2003**, *210*, 84.
[42]  P. Ares, G. Lopez-Polin, C. Hermosa, F. Zamora, J. Gomez-Herrero, C. Gomez-Navarro, *2D Mater.* **2015**, *2*, 035008.
[43]  M. Jaafar, G. Lopez-Polin, C. Gomez-Navarro, J. Gomez-Herrero, *Appl. Phys. Lett.* **2012**, *101*, 263109.
[44]  G. Kresse, D. Joubert, *Phys. Rev. B* **1999**, *59*, 1758.
[45]  a) G. Kresse, J. Furthmuller, *Comput. Mater. Sci.* **1996**, *6*, 15; b) G. Kresse, J. Furthmuller, *Phys. Rev. B* **1996**, *54*, 11169; c) G. Kresse, J. Hafner, *Phys. Rev. B* **1993**, *47*, 558; d) G. Kresse, J. Hafner, *Phys. Rev. B* **1994**, *49*, 14251.
[46]  J. P. Perdew, K. Burke, M. Ernzerhof, *Phys. Rev. Lett.* **1996**, *77*, 3865.
[47]  A. Tkatchenko, M. Scheffler, *Phys. Rev. Lett.* **2009**, *102*, 073005
[48]  H. J. Monkhorst, J. D. Pack, *Phys. Rev. B* **1976**, *13*, 5188.
[49]  a) T. P. M. Goumans, A. Wander, W. A. Brown, C. R. A. Catlow, *Phys. Chem. Chem. Phys.* **2007**, *9*, 2146; b) G. M. Rignanese, A. De Vita, J. C. Charlier, X. Gonze, R. Car, *Phys. Rev. B* **2000**, *61*, 13250.
[50]  M. K. Abyaneh, L. Gregoratti, M. Amati, M. Dalmiglio, M. Kiskinova, *e-J. Surf. Sci. Nanotech.* **2011**, *9*, 158.
[51]  L. Gregoratti, A. Barinov, E. Benfatto, G. Cautero, C. Fava, P. Lacovig, D. Lonza, M. Kiskinova, R. Tommasini, S. Mahl, W. Heichler, *Rev. Sci. Instrum.* **2004**, *75*, 64.
[52]  M. Schmid, H.-P. Steinrueck, J. M. Gottfried, *Surf. Interface Anal.* **2014**, *46*, 505.

Supporting Information

**Tunable Graphene Electronics with Local Ultrahigh Pressure**

P. Ares[1]*, M. Pisarra[2,3], P. Segovia[1,4], C. Díaz[2,4,5], F. Martín[2,3,4], E. G. Michel[1,4], F. Zamora[3,4,5,6], C. Gómez-Navarro[1,4], and J. Gómez-Herrero[1,4]*

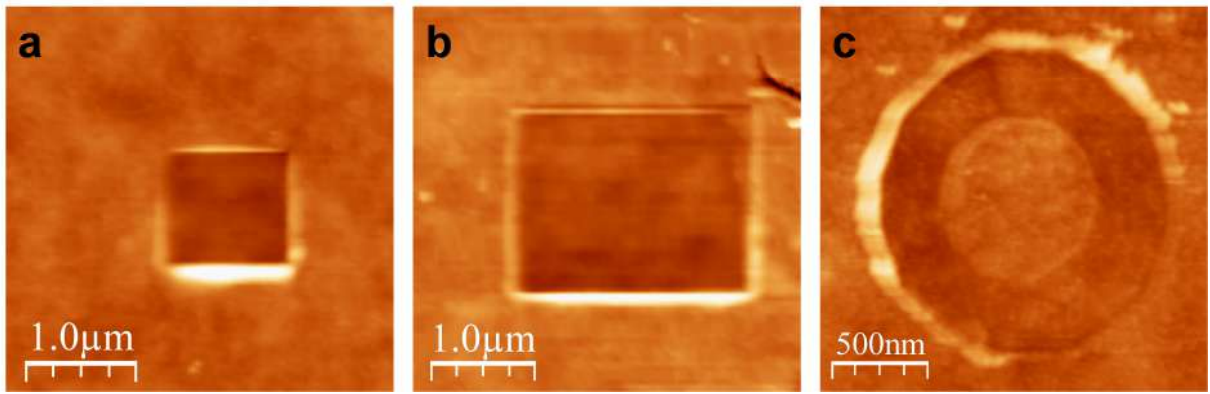

**Figure S1.** AFM topographic images of different graphene areas modified upon ultrahigh pressure. a) 1x1 µm$^2$ square area. b) 2x1.5 µm$^2$ rectangular area. c) Ring area with external diameter 1.5 µm and internal diameter 0.8 µm. Z scale 11 nm for all the images.

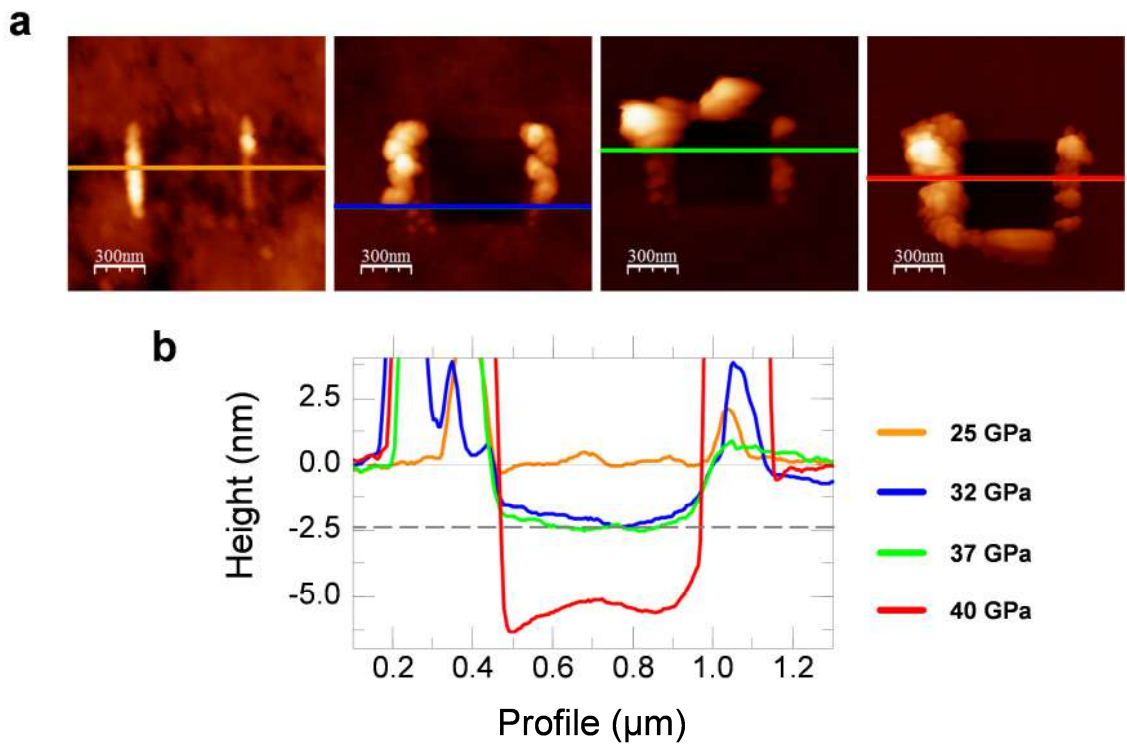

**Figure S2.** Ultrahigh pressure modifications on SiO$_2$. a) AFM topographic images showing 600 x 600 nm$^2$ areas modified under different pressures. b) Selected height profiles from the different modified regions. Horizontal dashed line marks the mean depth for pressures of 32 and 37 GPa.

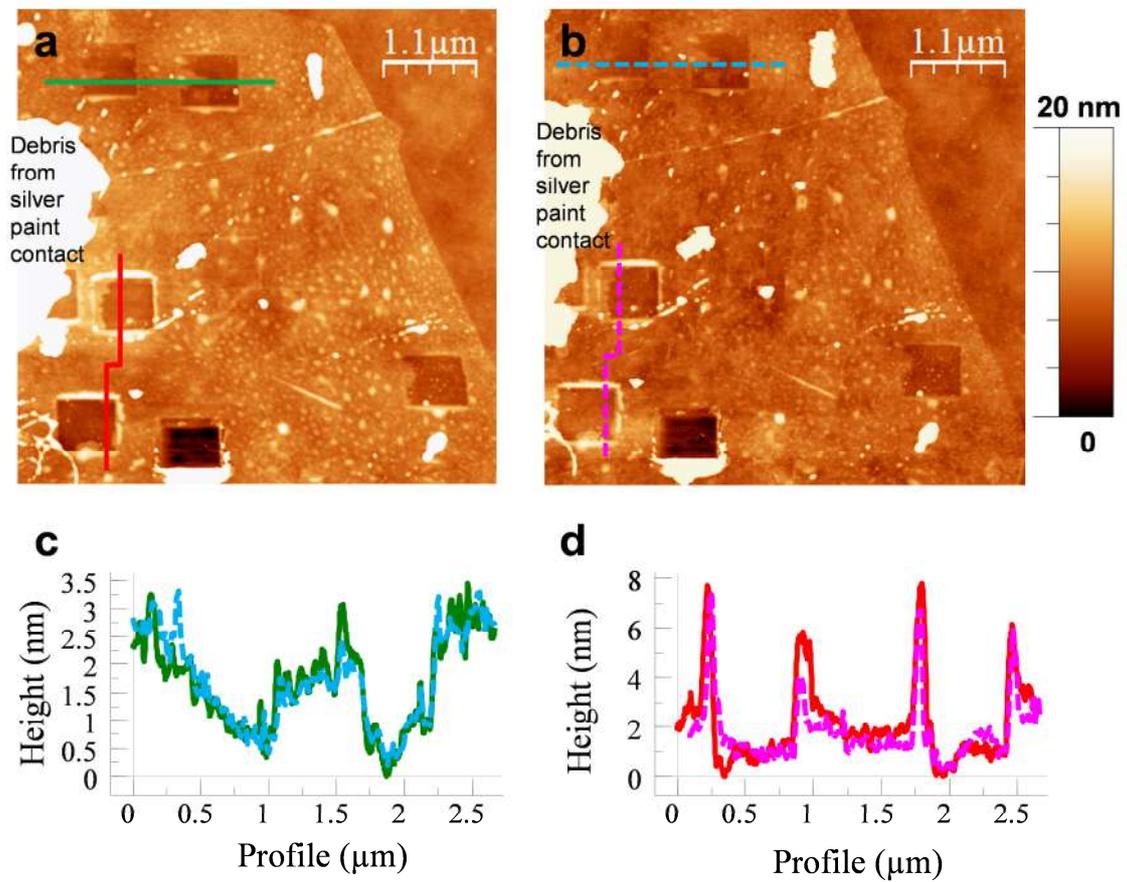

**Figure S3.** AFM topographic images of ultrahigh pressure modifications showing environmental stability. a) Image taken after modifications. b) Same as in a) but after four months storing the sample in ambient conditions. c,d) Height profiles along the lines in a) and b).

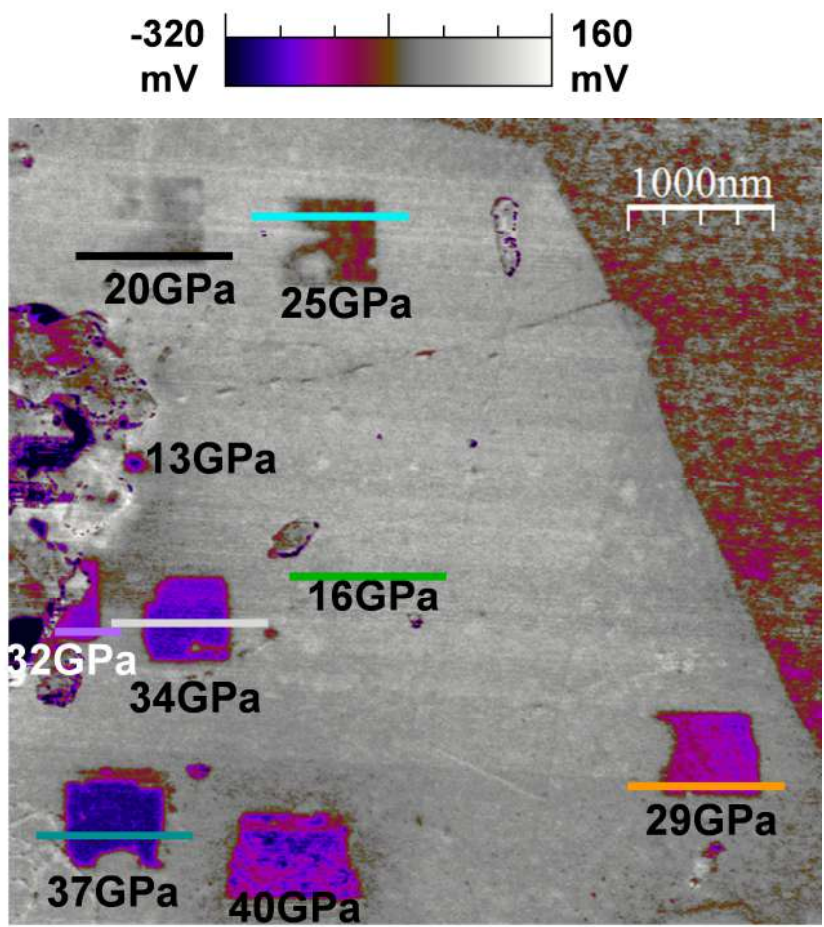

**Figure S4**. KPFM image of the modified regions. Note the area modified at 32 GPa is partially covered by a silver paint nano-drop deposited accidentally when placing an electrode for potential reference. The horizontal lines correspond to the Contact Potential Difference profiles shown in Figure 2b. The stable CPD value for the unmodified graphene over the whole image allows us to assume that the work function of the tip did not vary during the KPFM data acquisition.

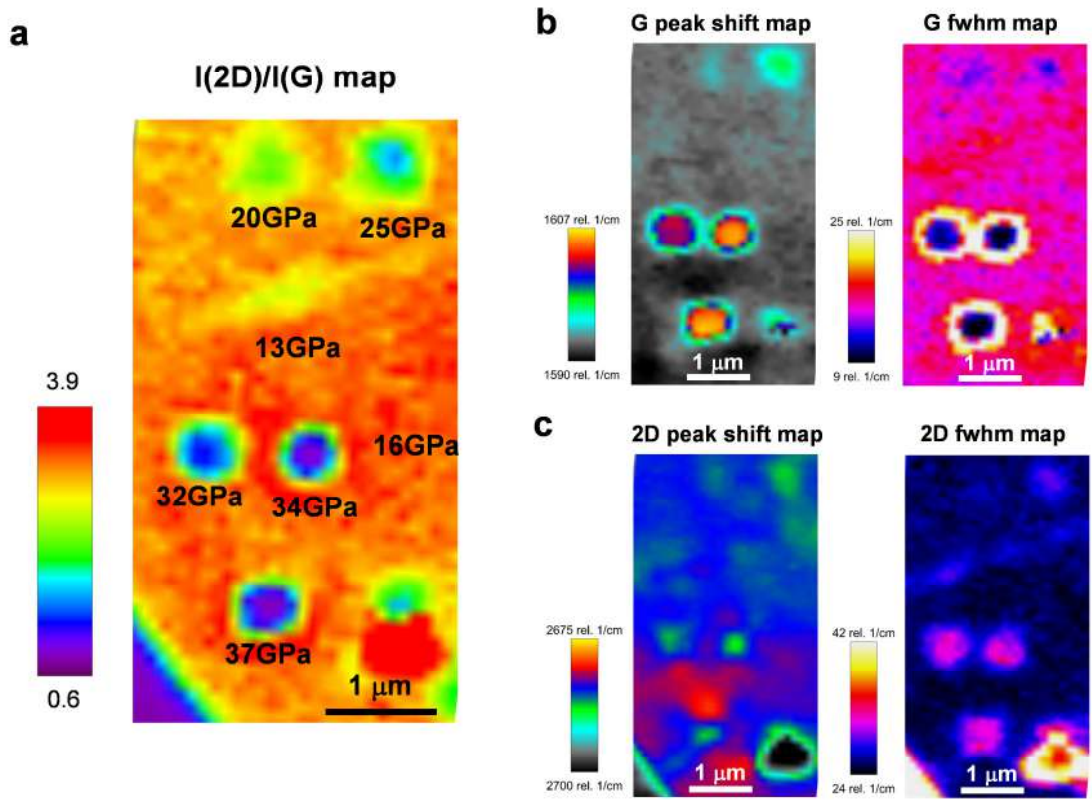

**Figure S5.** Raman mapping of different magnitudes of ultrahigh pressure modified areas. a) I(2D)/I(G) map. b) G peak. c) 2D peak. Left panels: peak position shift. Right panels: full width at half maximum (fwhm) variations.

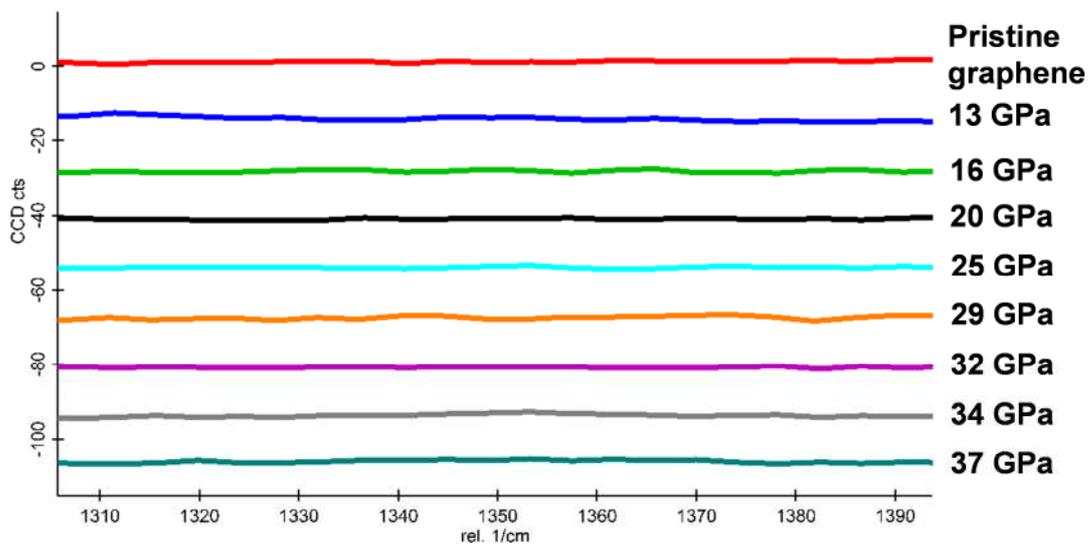

**Figure S6.** Raman spectra around 1350 cm$^{-1}$. No D peak is observed at any pressure.

**Charge effects in SPEM data**

Although graphene flakes were always electrically grounded, charge effects produced by the highly focused beam were observed in some cases, probably due to a poor contact with the ground electrode. In these cases, the C 1s peak position changed as a function of beam exposure, even for successive accumulation scans on the same spot. Data shown in this paper do not present charge effects.

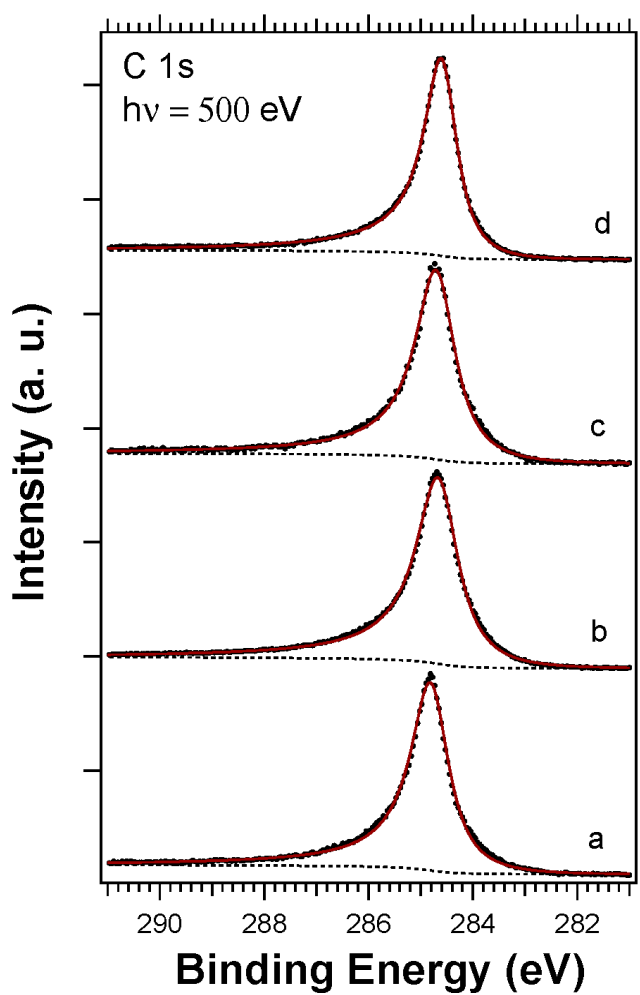

**Figure S7.** Representative C 1s peaks from pristine single-layer flakes from different samples. The spectra are vertically shifted for the sake of clarity. Dots are experimental points, red lines are fits to the data (see Experimental section), and the dotted lines are the Shirley background. A single component is used in all cases with $\alpha = 0.14$ and FWHM of a) 0.98 eV, b) 1.07 eV, c) 1.02 eV and d) 0.88 eV.

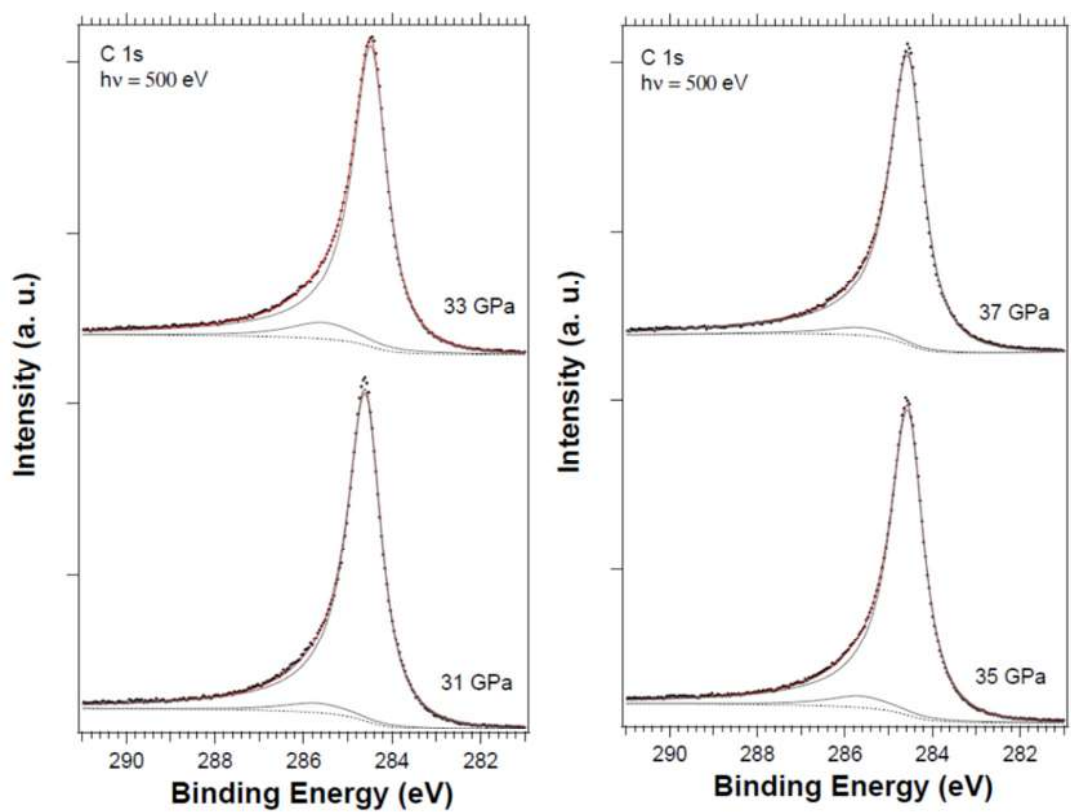

**Figure S8.** Left and right panels: deconvolution of representative C 1s peaks as a function of the applied pressure. The spectra are vertically shifted in each panel for the sake of clarity. Dots are experimental points, red lines are fits to the data (see Experimental section), the dotted lines are the Shirley background and the grey lines are the two components used.

**Table S1**: Summary of binding energies (BE) and intensity for the two components observed in the C 1s peak as a function of applied pressure.

| Applied pressure (GPa) | Component 1 | | Component 2 | |
| --- | --- | --- | --- | --- |
| | BE (eV) | Intensity | BE (eV) | Intensity |
| 0 (pristine) | 284.75 | 14499 | - | 0 |
| 31 | 284.65 | 14310 | 285.68 | 740 |
| 32 | 284.63 | 12318 | 285.65 | 1242 |
| 33 | 284.52 | 13167 | 285.55 | 1345 |
| 35 | 284.61 | 13495 | 285.64 | 933 |
| 37 | 284.61 | 12820 | 285.64 | 641 |

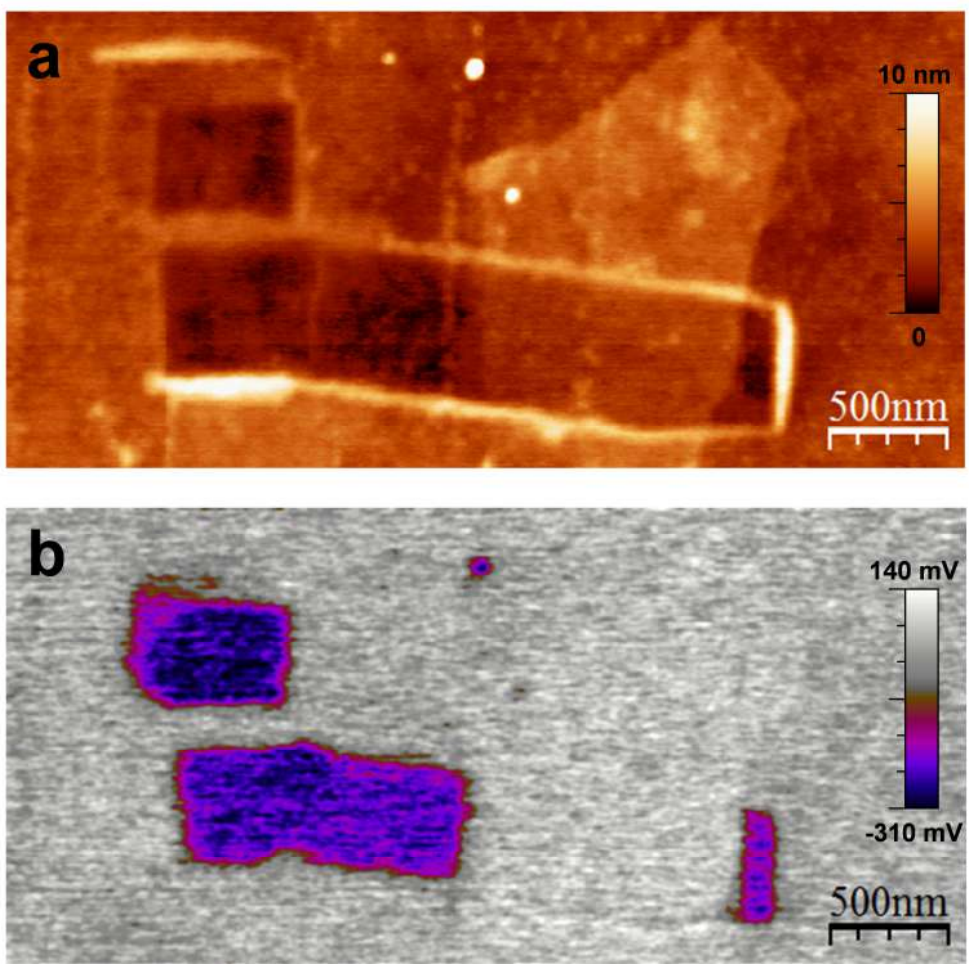

**Figure S9.** a) Topography and b) KPFM images of ultrahigh pressure modifications carried out in single and few layer graphene regions. Whereas in the single layer modified areas a clear decrease in the CPD is observed (representative of a p-doping effect), in the modified area of the few layer graphene region no changes in the CPD are detected.

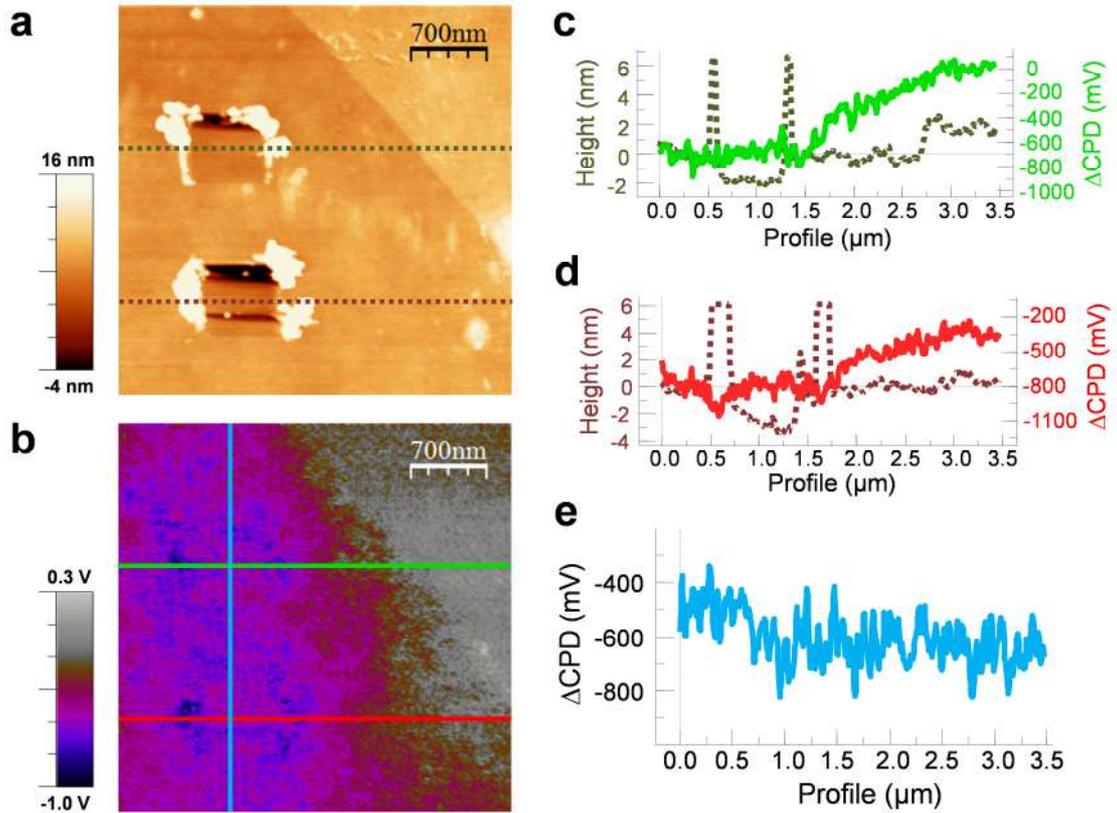

**Figure S10.** a) Topography and b) KPFM images of $SiO_2$ modified regions under 30 GPa (top area) and 38 GPa (bottom area). c,d) Topography (dotted) and KPFM (solid) profiles corresponding to the horizontal lines in a,b). e) KPFM profile along the vertical line in b). A negligible change in the $SiO_2$ electronic configuration due to the effect of tip induced strain is observed.

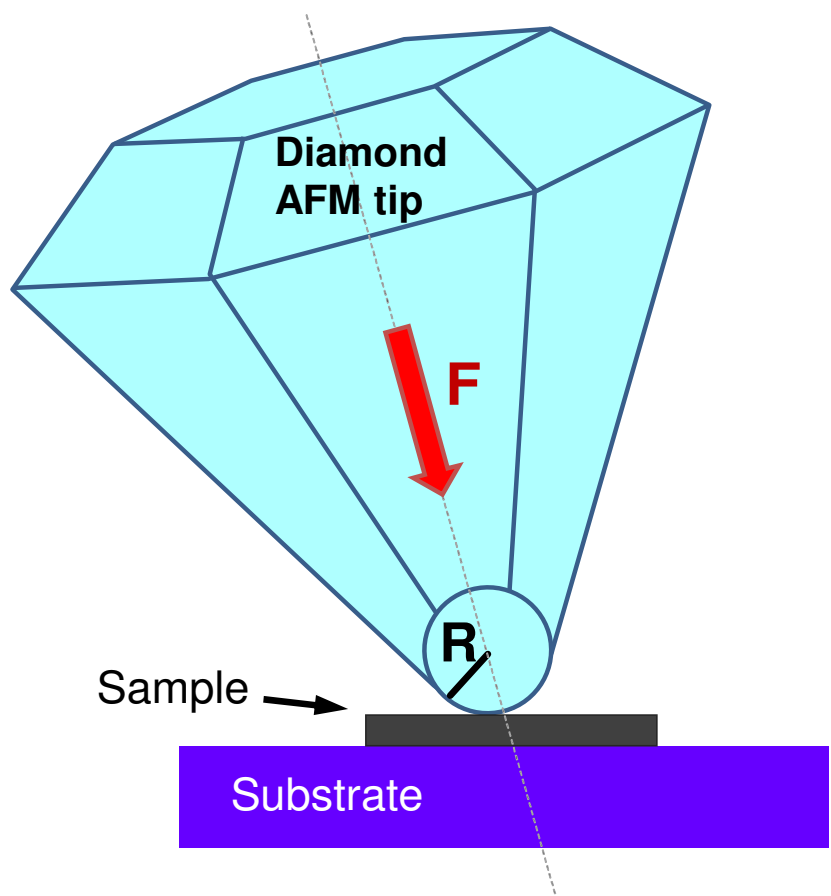

**Figure S11.** Schematics of a nano-anvil cell by using a diamond AFM tip.

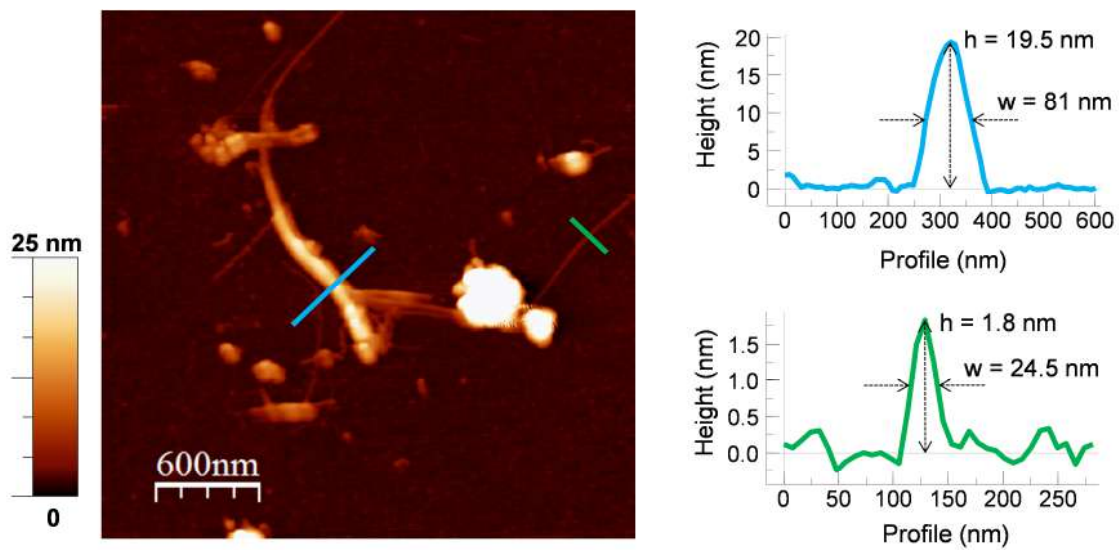

**Figure S12.** The tip radius has been estimated from the topographic images of carbon nanotubes of different heights using the expression: $R = \frac{w^2}{8h}$, where *R* is the tip radius and *w* is the apparent width measured for a carbon nanotube of height *h* ([P. Markiewicz and M. C. Goh, Langmuir 10, 5 (1994)]).

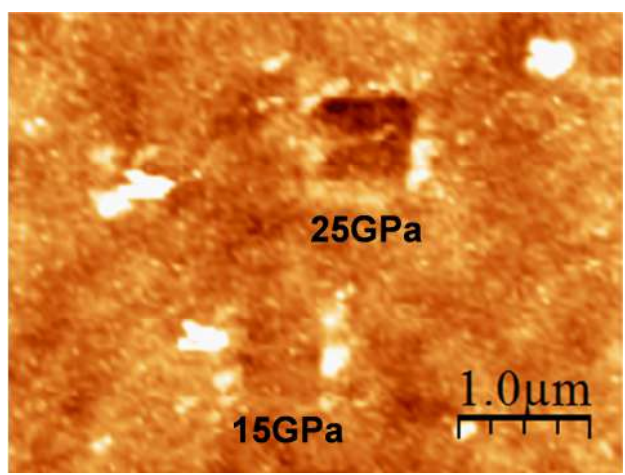

**Figure S13.** AFM topographic areas modified using high-density diamond like carbon spherical tips B50 from Nanotools with a nominal force constant of 40 N m$^{-1}$ and a nominal tip radius of 50 (± 5) nm. Z scale 3.5 nm.

**DFT Calculations**

**Modelling of the SiO$_2$ surface.** To model the SiO$_2$ surface in the DFT calculations we chose the (001) surface of alpha-quartz, which has hexagonal symmetry. Due to the particular atom arrangement, 3 different surface terminations are possible, which in general are rather reactive if the dangling bonds are not saturated (Figure S14). We also analyzed another rather stable surface, which is the result of a surface reconstruction. This surface termination is called "Superdense" and does not present dangling bonds.

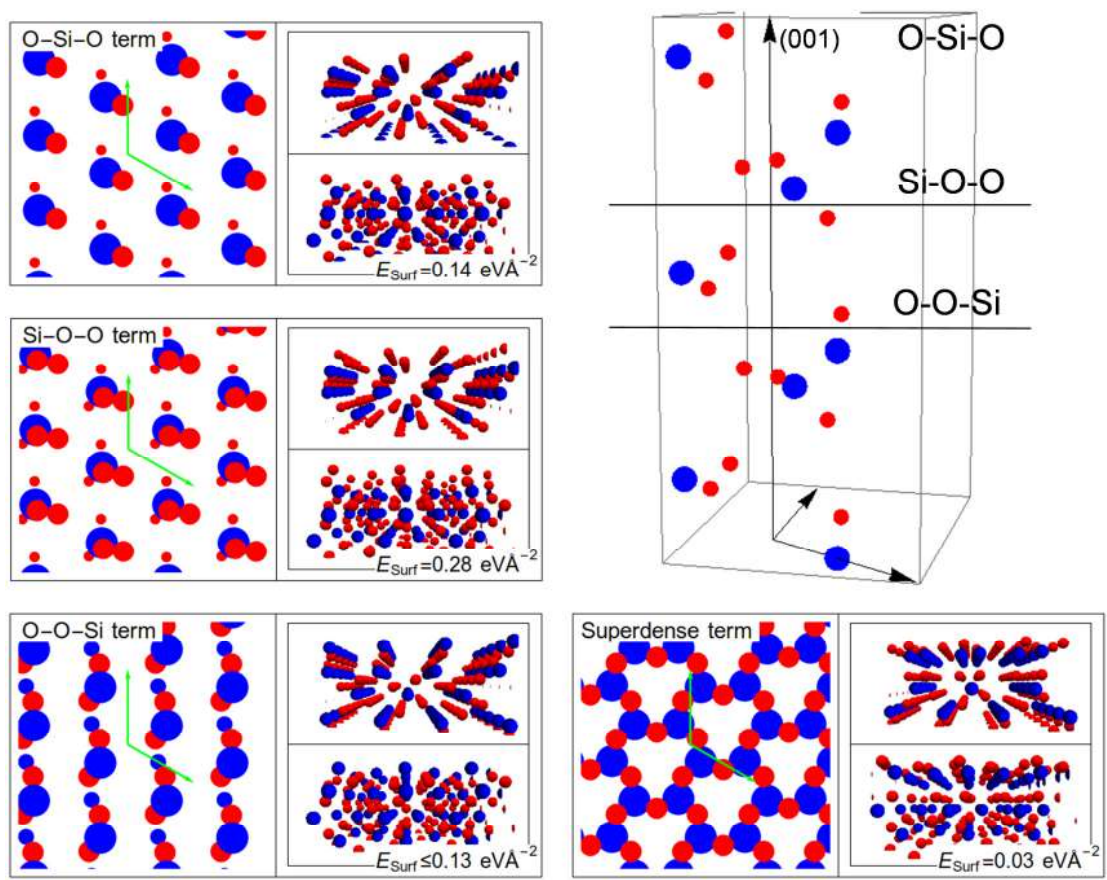

**Figure S14**. Hard Ball model illustrating the possible surface terminations for the (001) α-quartz geometry; in the "Superdense" termination the topmost 3 O layers and 2 Si layer are reconstructed. The color code is as follows: Red: O atoms; Blue: Si atoms. For clarity, in the top-views only the first few atomic layers are shown; bigger symbols represent atoms of the topmost layer. For each of the geometries, in the bottom right corner, an estimation of the surface energy is given: high values correspond to highly reacting surfaces.

**Graphene Adsorption Study.** Regarding the adsorption of graphene on SiO$_2$, it is worth mentioning that, due to the complexity of the experimental system under study (amorphous SiO$_2$, high roughness of the surface, ambient conditions of the experiments…), a faithful ab initio simulation is not feasible. For this reason, we ran hundreds of unconventional DFT based geometry optimizations, in which we use *extreme* starting positions for the four surface terminations shown in Figure S14. A non-perfect graphene layer (we added random displacements to each C atoms) was placed very close to the SiO$_2$ surface, at a distance of less than 1 Å. We deliberately chose not to saturate the dangling bonds of the surface. In this way we took into account the effect of the high pressure on both the graphene layer, by means of the random displacements, and on the surface, which presented high reactive centers. Most of the geometry optimizations ended up with the physisorption of graphene. In several cases we found local minima, in the configuration space, in which some of the C atoms formed covalent bonds with the surface atoms; in most of them, the final graphene layer was heavily damaged. Among the others, we found four different covalently bonded configurations in which the carbon atoms were organized into a distorted honeycomb lattice of graphene (see Figure S15 for the geometry details). To the best of our knowledge, at least two of them (the chemisorbed configurations for the O-Si-O and Superdense terminations) were not previously reported in the literature. For the O-Si-O termination, the chemisorbed minimum is characterized by a slightly buckled graphene layer, in which two C atoms (third neighbors) form covalent bonds with the surface O and Si atoms. In the Superdense chemisorbed minimum, a single covalent C-O bond is formed between one of the surface O atoms, which is displaced upwards, and one of the C atoms of graphene, which is displaced downwards. Due to this, the graphene lattice is distorted, but the honeycomb arrangement is still present. In the case of the Si-O-O termination, two chemisorbed minima have been found and are characterized by the formation of two C-O bonds between the two topmost O atoms of the surface and two C atoms of graphene, which are either first neighbors (1N) or third neighbors (3N). As shown in Figure 4 of the main text, these chemisorbed minima are connected to the physisorbed ones by pressure barriers of the order of 10-20 GPa. Thus, theoretical simulations confirm that the pressure used in the experiments is high enough to induce the formation of chemical bonds between the graphene layer and the $SiO_2$ substrate, provided that some high reacting centers of the surface are exposed during the process, a fact not unlikely under our experimental conditions.

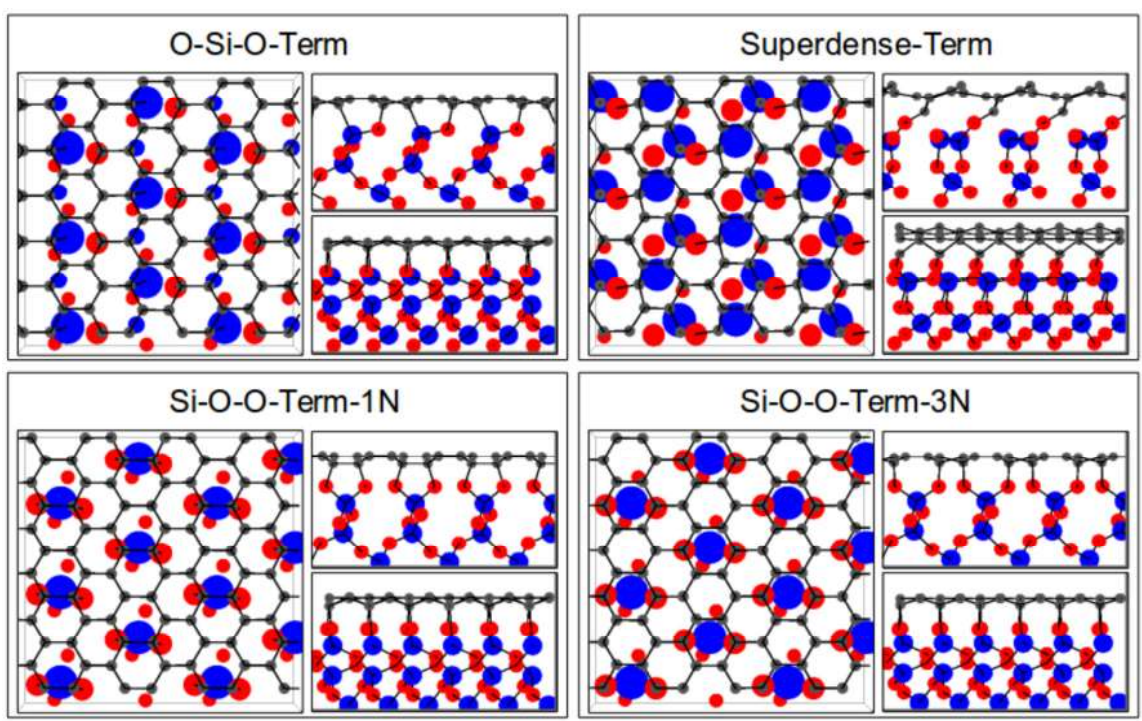

**Figure S15**. Covalently bonded configurations: Hard Ball model illustrating the covalently bonded configurations discussed in the main text. For each of the geometries we give the top view and two side views. The color code is as follows: Gray: C atoms; Red: O atoms; Blue: Si atoms. For clarity, in the top-views only the first few atomic layers are shown; bigger symbols represent atoms of the topmost layers.

**Supercell Calculations.** In the supercell calculations, we chose the starting positions as follows: in a small fraction of the supercell (a (1×1) $SiO_2$ unit cell) the graphene and surface coordinates were chosen so that the position of Figure S15 was reproduced; in the rest of the cell the dangling bonds of the substrate were saturated by H atoms, whereas the coordinates of the graphene atoms were gradually set to higher distances from the surface. As in the normal adsorption study, the system was then allowed to relax to the equilibrium position. Several geometry optimizations of this type were repeated, using the four different chemisorbed configurations of Figure S15, and changing the initial starting position for the C atoms that did not have chemical bonds. In most cases the geometry optimization ended in an equilibrium configuration in which no C-surface covalent bonds were present. This is mostly due to the fact that in the starting position, the C atoms that are not chemically bonded to the surface are on average repelled by the surface. The surface can be considered rather stable, because almost all the dangling bonds are saturated. As a result, the configuration in which graphene is flat far away from the surface (and the covalent bond disappears) is more energetically stable. However, two covalently bonded configurations survived this supercell test, namely the "Si-O-O-1N" and the "Superdense". In Figure 4 of the main text and Figure S16, we collect the results for the "Si-O-O-1N" geometry. Figure S17 shows similar results for the "Superdense" case. Observing that the (4×4) supercell has 128 C atoms, we see that the covalent contact of graphene with the surface is present if ~1% of the C atoms is covalently bonded to the surface (1.6% for the "Si-O-O-1N" geometry, in which two C-O bonds are present, and 0.7% for the "Si-O-O-1N" geometry where only one C-O bond is present). Regarding the $sp^2/sp^3$ ratio, we can argue that the C atoms that present some $sp^3$ component are the ones entering the first two groups discussed in the main text. In this way, we obtain an $sp^2/sp^3$ ratio of 4.7% for the "Si-O-O-1N" geometry and 2.3% for the "Superdense" geometry. The fact that, in the (4×4) supercell, the faraway atoms almost reached the z position of the physisorbed graphene implies that constructing a bigger supercell starting from the (4×4) would not induce further stress in the C-surface bond during the geometry optimization, implying that the chemical bond would also be found in the (5×5) and (6×6) supercell. For this reason, the figures given for the minimum percentage of covalent bonds that keep graphene strongly coupled to the surface and the $sp^2/sp^3$ ratio can be considered as upper bounds.

As a final remark, we observe that, even though DFT do not account for all the effects that can induce an effective doping of the graphene flake, we can use it to estimate the surface graphene charge transfer. No charge transfer takes place when graphene is stabilized only by van der Waals forces, as expected for an insulating surface. Interestingly, in the supercell calculation, when a low density of covalent bonds is present, we found that the far away areas present a p-doping compatible with a Fermi level shift of the order of -0.2 eV.

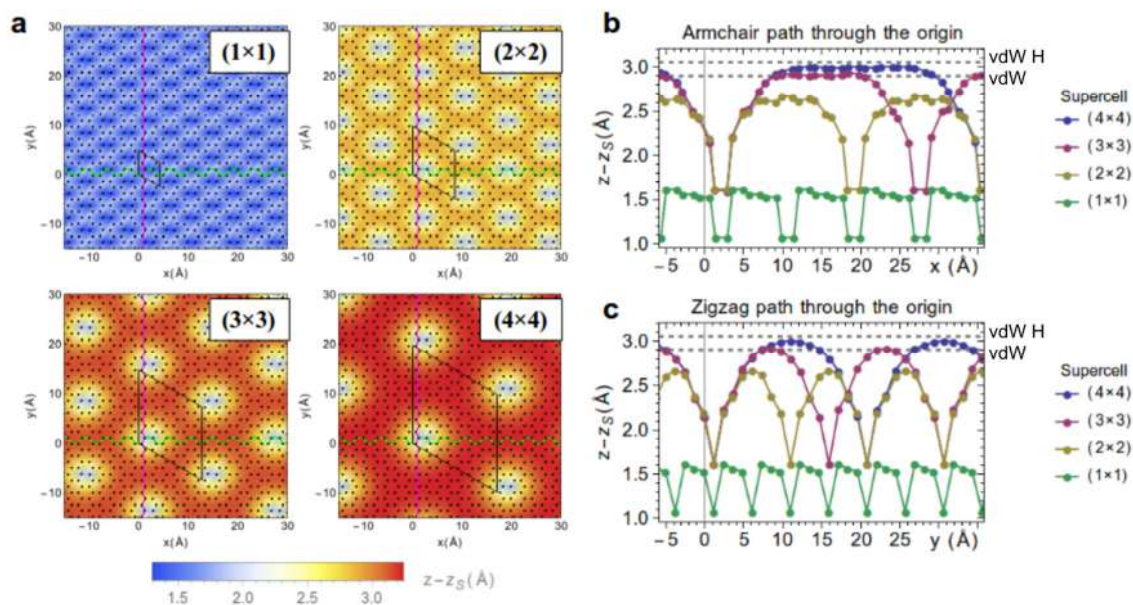

**Figure S16**. Complement of Figure 4 of the main text. a) z-value for the graphene atoms after the geometry optimization given as density plots. The same color code for all four supercells has been used to ease comparison. In each panel the black line delimits the unit cell, whereas the green and magenta lines mark the paths used for the corrugation plots in (b) and (c), respectively. b) Corrugation of the C atoms along an armchair path passing through the origin (green lines in (a)); points represent the actual position of the C atoms, while the lines are a guide for the eye. The horizontal gridlines mark the average position of the C atoms in the van der Waals attraction configuration without H atoms (vdW) or for the fully saturated surface (vdW H). c) Same as (b), now for a zigzag path passing through the origin (magenta lines in (a)).

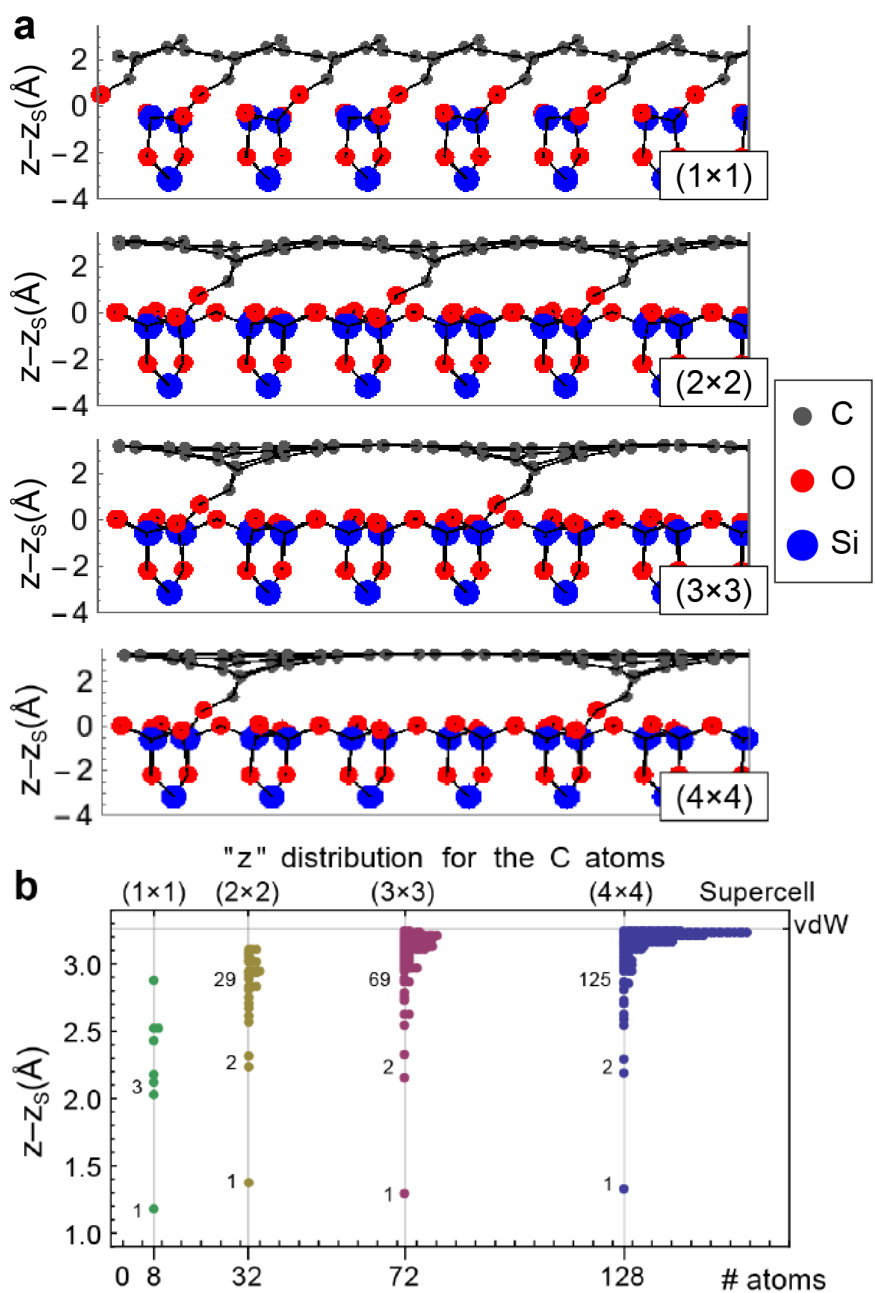

**Figure S17.** Supercell calculations for the "Superdense" termination. a) Side view after the geometry optimization for the (1×1), (2×2), (3×3), (4×4) supercells. b) z value distribution of the C atoms after the geometry optimization. The horizontal grid-lines mark the average z position of the C atoms in the van der Waals attraction configuration (vdW); the numbers inside the plot indicate the total number of C atoms in the corresponding group of histograms.